\documentclass[reprint,
superscriptaddress,
amsmath,amssymb,
aps,
prb,
]{revtex4-1}

\usepackage{graphicx}
\usepackage{comment}
\usepackage{dcolumn}
\usepackage{color}
\usepackage{bm}
\usepackage{url}
\usepackage{braket}
\usepackage{amsfonts,amsmath,mathtools}
\usepackage{dsfont}
\usepackage{hyperref}
\usepackage[all]{hypcap}
\setcitestyle{numbers,square}

\begin{document}

\title{High-temperature coherent transport in the XXZ chain in the presence of an impurity}

\author{Marlon Brenes}
\email{Corresponding author: brenesnm@tcd.ie}
\affiliation{School of Physics, Trinity College Dublin, College Green, Dublin 2, Ireland}
\author{Eduardo Mascarenhas}
\affiliation{Department of Physics and SUPA, University of Strathclyde, Glasgow G4 0NG, Scotland, United Kingdom}
\author{Marcos Rigol}
\affiliation{Department of Physics, The Pennsylvania State University, University Park, PA 16802, USA}
\affiliation{Kavli Institute for Theoretical Physics, University of California, Santa Barbara, California 93106, USA}
\author{John Goold}
\affiliation{School of Physics, Trinity College Dublin, College Green, Dublin 2, Ireland}
\affiliation{Kavli Institute for Theoretical Physics, University of California, Santa Barbara, California 93106, USA}

\date{\today}

\begin{abstract}
We study high-temperature spin transport through an anisotropic spin-$\frac{1}{2}$ Heisenberg chain in which integrability is broken by a single impurity close to the center of the chain. For a finite impurity strength, the level spacing statistics of this model is known to be Wigner-Dyson. Our aim is to understand if this integrability breaking is manifested in the high-temperature spin transport. We focus first on the nonequilibrium steady state (NESS), where the chain is connected to spin baths that act as sources and sinks for spin excitations at the boundaries. Using a combination of open quantum system theory and matrix product operators techniques, we extract the transport properties by means of a finite-size scaling of the spin current in the NESS. Our results indicate that, despite of the formation of a partial domain wall in the steady state magnetization (and despite the Wigner-Dyson level spacing distribution of the model), transport remains ballistic. We contrast this behavior with the one produced by a staggered magnetic field in the XXZ chain, for which it is known that transport is diffusive. By performing a numerical computation of the real part of the spin conductivity, we show that our findings are consistent with linear response theory. We discuss subtleties associated with the apparent vanishing of the Drude in the presence of an impurity.
\end{abstract}

\maketitle

\section{Introduction}
\label{sec:intro}
A recurring question in the theory of dynamical systems and nonequilibrium statistical mechanics is: How does macroscopic hydrodynamic behavior emerge from the underlying microscopic physics? Even in the classical domain recovering macroscopic linear phenomenology such as Fick's law for the particle current and Fourier's law for the heat current is highly nontrivial. It is known that there are systems for which Hamiltonian dynamics do not lead to this macroscopic phenomenology, particularly when conservation laws are at play~\cite{zotos1997transport, benenti2013conservation}. The consensus is that one needs nonlinear interactions, which lead to chaos and, hence, to incoherent transport~\cite{lepri2003thermal, dhar2008heat, chen2014nonintegrability}.

In quantum systems, understanding how this complexity emerges brings us to the domain of quantum chaos~\cite{haake2013quantum,chirikov1997linear}. In the past two decades, interest in quantum chaotic behavior of many-body quantum systems has seen an unprecedented revival~\cite{Zelevinsky1996, Polkovnikov:2011, Yukalov:2011, Eisert:2015, Goold:2016, Gogolin:2016, Borgonovi:2016, Alessio:2016}. Strides in experimental ultracold atomic physics~\cite{Greiner:2002, Kinoshita:2006, Trotzky:2012, Kaufman2016, tang2018thermalization} have lead to new lines of research, and have highlighted the role of integrability and its breaking in thermalization (or lack thereof)~\cite{Rigol:2008, Rigol:2009} and transport~\cite{vasseur2016nonequilibrium}.

The connection between thermalization and integrability breaking has been extensively studied theoretically~\cite{Zelevinsky1996, Polkovnikov:2011, Yukalov:2011, Eisert:2015, Goold:2016, Gogolin:2016, Borgonovi:2016, Alessio:2016} and, recently, a beautiful experiment was performed in which it was possible to tune the integrability breaking of a low-dimensional gas of dipolar atoms~\cite{tang2018thermalization} and, that way, tune the relaxation rates to thermal equilibrium values. In terms of transport, the emergence of hydrodynamics due to integrability breaking in quantum many-body systems is far less understood.

From the theoretical perspective, studying transport in nonintegrable models represents a significant computational challenge, as both large system sizes and long-time limits are required~\cite{Heidrich-Meisner_Honecker_review_07}. This requirement is even more prevalent at high energies where effective low-energy field theories fail~\cite{prelovvsek2002transport}. A relatively modern approach for extracting high-temperature transport properties of nonintegrable one-dimensional quantum systems is known as boundary driving~\cite{karevski2009quantum, MarkoStaggered1, Benenti:2009, MarkoStaggered1, Znidaric:2010, Znidaric:2010b, Prosen:2011, ZnidaricXXZspintransport, Mendoza:2013a, Mendoza:2013b, karevski2013exact, Landi:2015}. Boundary driving is a setup which stems from the theory of open quantum systems, in which Lindblad jump operators are applied at the boundaries of the chain in order to model spin sources and sinks that drive the chain into a nonequilibrium steady state (NESS). In some cases, it may be combined with the power of matrix product operator techniques~\cite{Schollwock2011} to reach system sizes beyond those accessible via full exact diagonalization or Lanczos based techniques. Transport properties can be determined by means of finite-size scaling of the current operator in the NESS. This approach has been successful in providing an accurate numerical characterization of high temperature transport properties of the XXZ model~\cite{Znidaric:2010, ZnidaricXXZspintransport} and of the ergodic regime of spin chains that exhibit many-body localization~\cite{vznidarivc2016diffusive, vznidarivc2017dephasing, mendoza2018asymmetry, schulz2018energy}. These works have shown that strong integrability breaking need not result in diffusive transport in the steady state, and that anomalous diffusion is ubiquitous.

Here, we focus on high-temperature transport in the spin-$\frac{1}{2}$ XXZ chain in the presence of integrability breaking in the form of a single (static) magnetic defect. This model is known in the literature to lead to quantum chaos~\cite{Santos:2004, santos2011domain, torres2014local, XotosIncoherentSIXXZ}. We contrast the results for that model with those from a model in which the (global) integrability breaking perturbation applied to the XXZ chain is a staggered magnetic field. The latter perturbation is known to render transport fully diffusive~\cite{MarkoStaggered1, Huang2013, juan2014a, BrenigTypicality2015}.

Our paper is structured as follows: In Sec.~\ref{sec:xxz_int}, we introduce the models and discuss their level spacing statistics. In Sec.~\ref{sec:config}, we review the boundary driving protocol, the basic ingredients of the theory of spin transport and finite-size scaling, and briefly describe the techniques used to obtain the solution to the steady state. The open system results are reported in Sec.~\ref{sec:results}. In Sec.~\ref{sec:linearresponse}, we present a closed system analysis based on Kubo's linear response theory. A summary of our results and an outlook are provided in Sec.~\ref{sec:conclusions}.

\section{XXZ model with integrability breaking}
\label{sec:xxz_int}

Our unperturbed Hamiltonian is the anisotropic spin-$\frac{1}{2}$ Heisenberg chain, also known as the spin-$\frac{1}{2}$ XXZ chain, which can be written as ($\hbar = 1$):
\begin{align}
\label{eq:h_xxz}
\hat{H}_{\textrm{XXZ}} = \sum_{i}\left[\alpha\left(\hat{\sigma}^x_{i}\hat{\sigma}^x_{i+1} + \hat{\sigma}^y_{i}\hat{\sigma}^y_{i+1}\right) + \Delta\,\hat{\sigma}^z_{i}\hat{\sigma}^z_{i+1}\right],
\end{align} 
where $\hat{\sigma}^\nu_{i}$, $\nu = x,y,z$, correspond to Pauli matrices in the $\nu$ direction at site $i$ in a one-dimensional lattice with $N$ sites. Boundary conditions are specified as {\em open} if the sum in Eq.~\eqref{eq:h_xxz} includes all the sites but the last one ($N-1$) and {\em periodic} if it includes all the sites ($N$). $\Delta$ is known as the anisotropy parameter [for $\Delta=1$, Hamiltonian~\eqref{eq:h_xxz} is the Hamiltonian of the spin-$\frac{1}{2}$ Heisenberg chain]. The spin-$\frac{1}{2}$ XXZ chain is integrable and exactly solvable via Bethe ansatz~\cite{ShastryBethe1990, Cazalilla:2011}. In what follows, we only consider $\alpha = 1$ and $0 < \Delta < 1$. In order to break integrability, we use the following modifications to $\hat{H}_{\textrm{XXZ}}$ in Eq.~\eqref{eq:h_xxz}:
\begin{align}
\label{eq:h_si}
\hat{H}_{\textrm{SI}} &= \hat{H}_{\textrm{XXZ}} + h\, \hat{\sigma}^z_{N/2}\,,\\
\label{eq:h_sf}
\hat{H}_{\textrm{SF}} &= \hat{H}_{\textrm{XXZ}} + b\,\sum_{i\,odd} \hat{\sigma}^z_{i}\,.
\end{align}
In Eq.~\eqref{eq:h_si}, we introduce a single magnetic impurity in one of the sites about the center of the chain. We consider cases in which $N$ is even, and introduce the defect at site $i = N / 2$. In Eq.~\eqref{eq:h_sf}, we introduce a global staggered transverse field (see Fig.~\ref{fig:diagram}). We refer to the former as the {\em single impurity} model and to the latter as the {\em staggered field} model. These models commute with the total magnetization operator in the $z$ direction, $[\hat{H}_{\textrm{SI}}, \sum_i\hat{\sigma}^z_i] = [\hat{H}_{\textrm{SF}}, \sum_i\hat{\sigma}^z_i]=0$. Only the magnitude of $h$ and $b$ matter for the results we discuss in what follows.

\subsection{Level spacing statistics}
\label{sec:lss}

The distribution $P(s_n)$ of spacings $s_n$ of neighboring energy levels shows different behavior depending on whether a quantum system is chaotic or integrable, and they are often employed as a diagnostic tool~\cite{Alessio:2016}. For an integrable system, energy levels are expected to be independent from each other and crossings are not prohibited from occurring. Therefore, the statistics of the levels in this case is Poissonian,
\begin{align}
\label{eqn:poisson}
P(s) = e^{-s}.
\end{align}

On the other hand, a hallmark of quantum chaos is that energy levels repel each other and become correlated. As obtained from random matrix theory, the level spacings of quantum chaotic systems with time-reversal invariance exhibit a Wigner-Dyson distribution given by 
\begin{align}
\label{eq:wddistro}
P(s) = \frac{\pi s}{2}e^{-\frac{\pi s^2}{4}}.
\end{align}

\begin{figure}[!t]
\fontsize{13}{10}\selectfont 
\centering
\includegraphics[width=0.9\columnwidth]{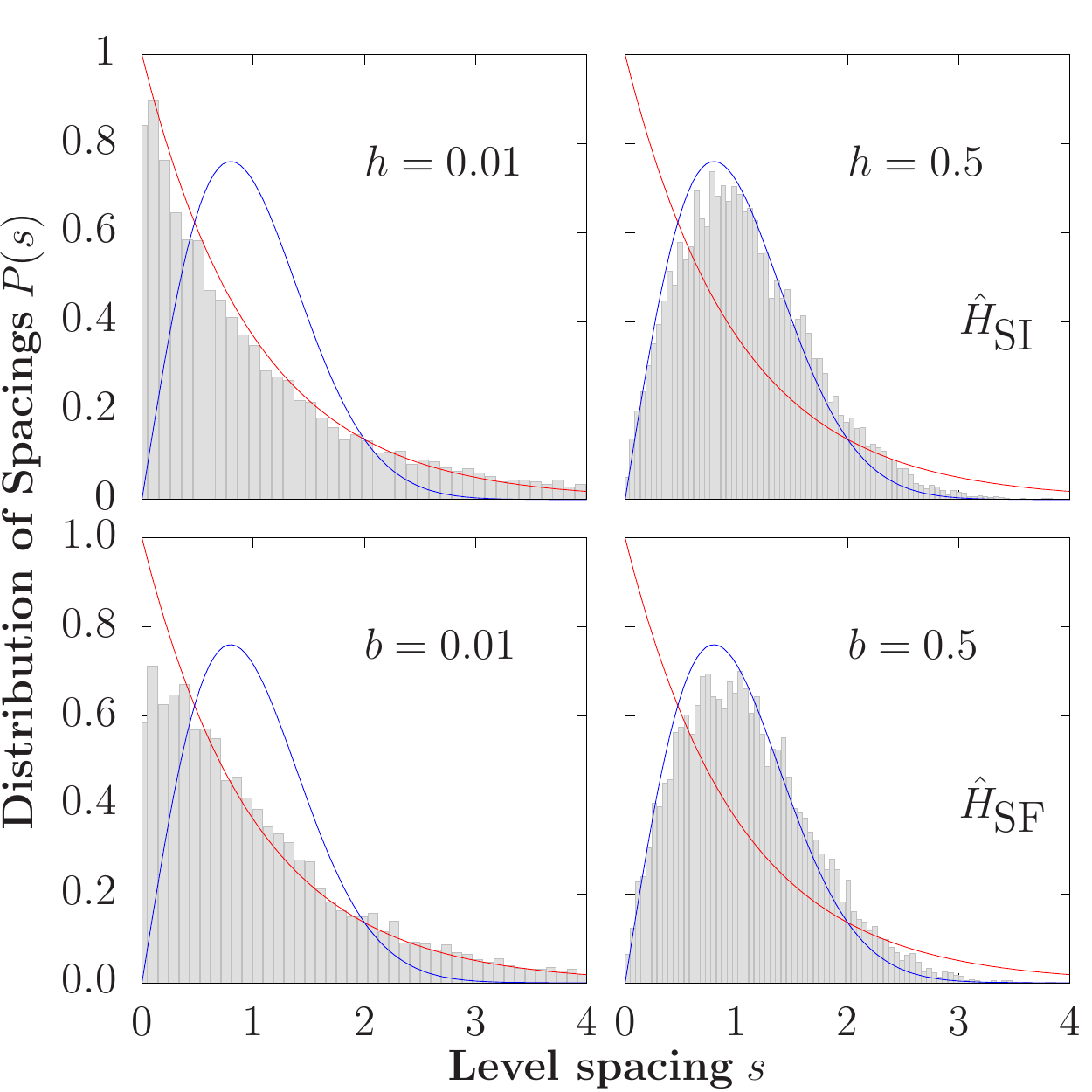}
\caption{Level spacing distribution $P(s)$ for the anisotropic Heisenberg model (top) in the presence of a single magnetic impurity [see Eq.~\eqref{eq:h_si}], and (bottom) in the presence of a staggered magnetic field [see Eq.~\eqref{eq:h_sf}]. The red line corresponds to a Poisson distribution [Eq.~\eqref{eqn:poisson}], while the blue line depicts a Wigner-Dyson distribution [Eq.~\eqref{eq:wddistro}]. The results shown are for chains with open boundary conditions, $N = 16$, $\Delta = 0.5$, $\sum_{j=1}^{N} \langle \hat{\sigma}^z_j \rangle = 0$, and two values of $h$ and $b$.}
\label{fig:lss}
\end{figure}

In Fig.~\ref{fig:lss}, we show the behavior of the distribution $P(s_n)$ for both the $\hat{H}_{\textrm{SI}}$ model in Eq.~\eqref{eq:h_si}, for different strengths of the impurity, and for the $\hat{H}_{\textrm{SF}}$ model in Eq.~\eqref{eq:h_sf}, for different strengths of the staggered field. The calculations were done in the zero magnetization sector, $\sum_{j=1}^{N} \langle \hat{\sigma}^z_j \rangle = 0$, in chains with $N=16$ sites and open boundary conditions. Our results confirm that, as previously observed for $\hat{H}_{\textrm{SI}}$~\cite{Santos:2004, santos2011domain, torres2014local, XotosIncoherentSIXXZ} and for $\hat{H}_{\textrm{SF}}$~\cite{juan2014a,Huang2013}, the level spacing distribution becomes Wigner-Dyson as one increases the magnitude of $h$ and $b$, respectively, without changing $\Delta$ and $N$. For the single impurity model, at fixed $\Delta$ and $N$, the probability distribution of energy spacings was shown in Ref.~\cite{XotosIncoherentSIXXZ} to be of the Wigner-Dyson type for a wide range of values of $h$. It was also shown there that, increasing $N$ at fixed $\Delta$ increases the range of values of $h$ for which quantum chaotic behavior occurs. As for systems in which integrability is broken by means of global perturbations~\cite{Santos:2010b, Santos:2010}, for $\Delta\neq0$ in the thermodynamic limit one expects quantum chaotic behavior to occur whenever $h\neq0$ and $h\neq\infty$.

In order to obtain the correct level spacing distribution, an {\em unfolding procedure} of the spectrum needs to be used in which one locally rescales the energies, so that the local density of states (LDOS) is normalized to 1. The symmetries of the model have to be taken into account as well, given that energy levels from different symmetry subsectors (subspaces of the Hilbert space) are independent from each other and therefore uncorrelated \cite{Santos:2010b, Santos:2010}. For $\hat{H}_{\textrm{SI}}$, the reflection symmetry of the XXZ model is broken by the impurity, while for $\hat{H}_{\textrm{SF}}$ there is a related remaining symmetry that needs to be resolved. The key point to be emphasized here is that both integrability breaking perturbations, a {\em local} one in $\hat{H}_{\textrm{SI}}$ and a {\em global} one in $\hat{H}_{\textrm{SF}}$, lead to the same quantum chaotic behavior of the level spacing distributions. In the rest of the paper, we focus on the transport properties of those quantum chaotic models.

\section{Nonequilibrium configuration for spin transport}
\label{sec:config}

\begin{figure}[!t]
\centering
\includegraphics[width=\columnwidth]{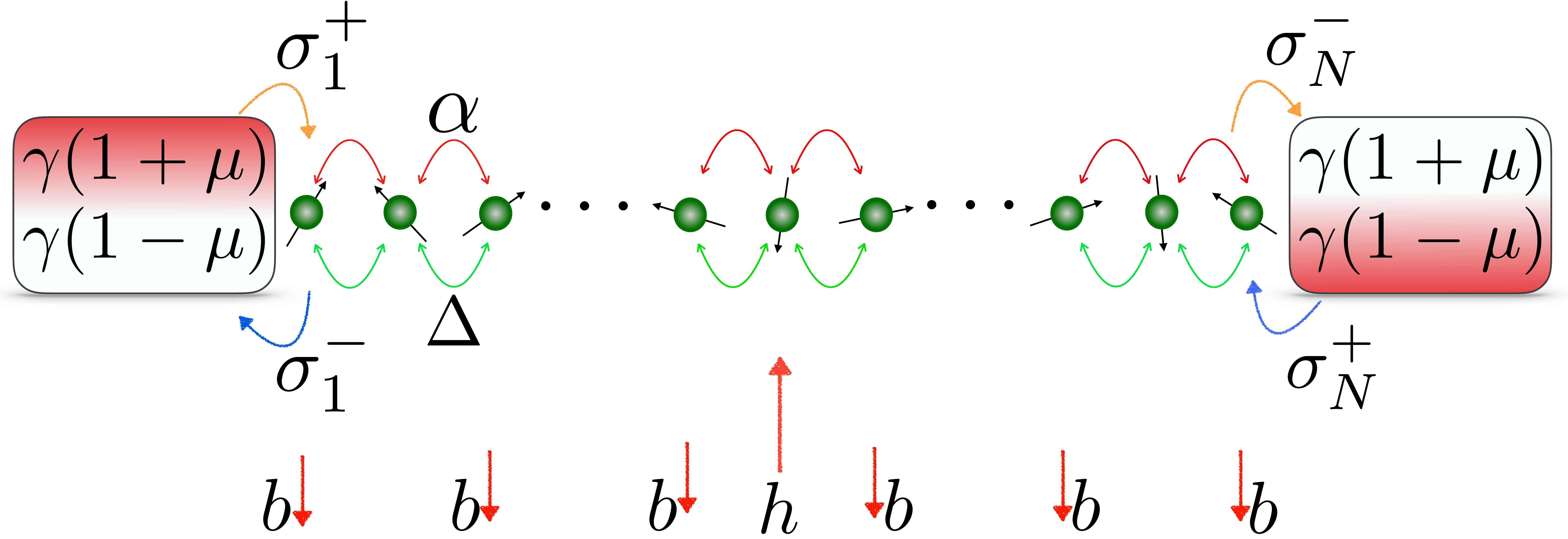}
\caption{Diagrammatic depiction of the nonequilibrium configuration used to study transport in the boundary-driven scheme. Excitations induced by the baths can propagate through the system because of the first two terms in Eq.~\eqref{eq:h_xxz}, top red arrows, while interactions occur because of the third term in Eq.~\eqref{eq:h_xxz}, bottom green arrows. The system-bath coupling strength is given by $\gamma$, while $\mu$ represents the driving strength. A sufficiently strong (but finite) field in either configuration (a single magnetic impurity of strength $h$ or a staggered field of strength $b$) renders the system nonintegrable.}
\label{fig:diagram}
\end{figure}

In order to study transport in a genuinely nonequilibrium steady-state in a long chain, we couple the latter to two Markovian baths that create and remove excitations at the boundaries. The dynamics of such a setup can be analyzed by means of the Lindblad master equation 
\begin{align}
\label{eq:lme}
\frac{d\hat{\rho}}{dt} &= -i[\hat{H},\hat{\rho}] + \mathcal{L}\{\hat{\rho}\} \nonumber \\ 
&= -i[\hat{H},\hat{\rho}] + \mathcal{L}_l\{\hat{\rho}\} + \mathcal{L}_r\{\hat{\rho}\},
\end{align}
where $\hat{\rho}$ is the density matrix of the system and $\mathcal{L}_{l,r}$ are dissipative superoperators that act on $\hat{\rho}$ inducing excitations in terms of spin creation and annihilation operators given by $\hat{\sigma}^{\pm}_j = (\hat{\sigma}^x_j \pm i\hat{\sigma}^y_j) / 2$ for site at position $j$. Specifically, we have
\begin{align}
\label{eq:lind1}
\mathcal{L}_m\{\hat{\rho}\} = \sum_{s=\pm} 2\hat{L}_{s,m}\,\hat{\rho}\, \hat{L}_{s,m}^{\dag} - \{\hat{L}_{s,m}^{\dag}\hat{L}_{s,m},\hat{\rho}\},
\end{align}
where $m = l, r$ and $\{\cdot\,,\cdot\}$ is the anticommutator. The operators in Eq.~\eqref{eq:lind1} are defined as follows:
\begin{align}
\label{eq:lind2}
\hat{L}_{+,l} = \sqrt{\gamma(1 + \mu)}\,\hat{\sigma}_1^{+}, \nonumber \\
\hat{L}_{-,l} = \sqrt{\gamma(1 - \mu)}\,\hat{\sigma}_1^{-}, \nonumber \\
\hat{L}_{+,r} = \sqrt{\gamma(1 - \mu)}\,\hat{\sigma}_N^{+}, \nonumber \\
\hat{L}_{-,r} = \sqrt{\gamma(1 + \mu)}\,\hat{\sigma}_N^{-},
\end{align}
where $\gamma$ is the bath coupling parameter and $\mu$ is a parameter that dictates the strength of the boundary driving. A diagrammatic depiction of the nonequilibrium configuration is presented in Fig.~\ref{fig:diagram}. The Lindblad master equation [Eq.~\eqref{eq:lme}] can be obtained from a microscopic derivation, such as the one used in the repeated interactions scheme, which allows one to obtain expressions for thermodynamic quantities such as heat and work~\cite{barra2015thermodynamic, PereiraHeatXXZ}.

\subsection{Spin current and steady state}
\label{sec:current}

The configuration described previously drives the system towards a nonequilibrium steady state, denoted by $\hat{\rho}_{\textrm{NESS}}$, given by
\begin{align}
\label{eq:supop}
\mathcal{W}\{\hat{\rho}_{\textrm{NESS}}\} =& -i[\hat{H},\hat{\rho}_{\textrm{NESS}}]  \nonumber \\ &+\mathcal{L}_l\{\hat{\rho}_{\textrm{NESS}}\} + \mathcal{L}_r\{\hat{\rho}_{\textrm{NESS}}\} = 0,
\end{align}
which implies that the steady state is the one that spans the null space of the superoperator $\mathcal{W}$. It can be proven that this state exists and is unique if and only if the set of operators $\{\hat{H}, \hat{L}_{+,l}, \hat{L}_{+,r}, \hat{L}_{-,l}, \hat{L}_{-,r}\}$ generate, under multiplication and addition, the entire Pauli algebra. This condition is fulfilled in our case~\cite{ProsenUnique}. Another property of the NESS is related to the time evolution of the system. Given the mathematical existence and uniqueness of this particular state, any initial state will converge to the NESS in the long time limit
\begin{align}
\label{eq:ness}
\lim_{t\rightarrow\infty}\hat{\rho}(t) = \hat{\rho}_{\textrm{NESS}}.
\end{align}

Since, by construction, we introduce an imbalance in the strength of the boundary driving $\mu$, the NESS is characterized by a constant flow of magnetization from one boundary to the other. The boundary driving parameter establishes the degree of imbalance between the Markovian baths and thus affects transport in the bulk of the spin chain. We focus on the regime $0 \leq \mu \leq 1$. For $\mu = 0$ there is no imbalance and the state in the bulk is given by an infinite temperature steady state, $\hat{\rho} = \mathds{1}/2^{N}$. For any nonzero $\mu$, effective spin excitations are introduced and removed from the system. For $\mu = 1$ the system is at maximum driving, i.e., maximum bias.    

We can determine the flux of magnetization by means of the equation dictating the dynamics of the expectation value of $\hat{\sigma}^z_i$. We then turn to Eq.~\eqref{eq:lme} to obtain, in the bulk of the chain
\begin{align}
\label{eq:magevo}
\frac{d\langle \hat{\sigma}^z_i \rangle}{dt} &= \frac{d}{dt}\textrm{Tr}\left(\hat{\rho}\hat{\sigma}^z_i\right) 
= \textrm{Tr}\left(\hat{\sigma}^z_i\frac{d\hat{\rho}}{dt}\right) 
= -i\,\textrm{Tr}\left(\hat{\sigma}^z_i[\hat{H},\hat{\rho}]\right) \nonumber \\
&= i\,\textrm{Tr}\left([\hat{H},\hat{\sigma}^z_i]\hat{\rho}\right);\quad \forall i = 2, \cdots, N-1\,.
\end{align}
Using Pauli matrix commutation relations, one obtains for Eq.~\eqref{eq:magevo}:
\begin{align}
\label{eq:magevocurrent}
\frac{d\langle \hat{\sigma}^z_i \rangle}{dt} = \langle \hat{j}_{i-1} \rangle - \langle \hat{j}_i \rangle;\quad \forall i = 2, \cdots, N-1,
\end{align}
where
\begin{align}
\label{eq:spincurrent}
\hat{j}_{i} \coloneqq 2\alpha_i\left(\hat{\sigma}^x_i\hat{\sigma}^y_{i+1} - \hat{\sigma}^y_i\hat{\sigma}^x_{i+1}\right).
\end{align}

We call this object the {\em spin current} operator. Up to this point, Eq.~\eqref{eq:magevocurrent} is ill-defined for the leftmost and the rightmost sites of the chain. However, we can obtain the dynamics of the magnetization in these sites by interpreting $\mu$ as the average magnetization of the Markovian baths, where we therefore identify 
\begin{align}
\label{eq:boundcurrent}
\frac{d\langle \hat{\sigma}^z_1 \rangle}{dt} &= \langle \hat{j}_{l} \rangle - \langle \hat{j}_1 \rangle, \\
\frac{d\langle \hat{\sigma}^z_N \rangle}{dt} &= \langle \hat{j}_{N-1} \rangle - \langle \hat{j}_r \rangle,
\end{align}
with the corresponding values of the current on the boundaries given by 
\begin{align}
\label{eq:boundarycurrent}
\langle \hat{j}_l \rangle &= \textrm{Tr}\left(\hat{\sigma}^z_1\mathcal{L}_l\{\hat{\rho}\}\right) = 4\gamma\,(\mu - \langle \hat{\sigma}^z_1 \rangle), \\
\langle \hat{j}_r \rangle &=  \textrm{Tr}\left(\hat{\sigma}^z_N\mathcal{L}_r\{\hat{\rho}\}\right) = 4\gamma\,(\mu + \langle \hat{\sigma}^z_N \rangle).
\end{align}

With these definitions, the continuity equation of the magnetization in the $z$ direction is consistent. In the NESS, the relation $d\langle \hat{\sigma}^z_i \rangle / dt= 0$ holds for all sites, which means that the spin current is homogeneous across the chain (in one dimension): 
\begin{align}
\label{eq:homogeneous}
\langle \hat{j}_l \rangle = \langle \hat{j}_1 \rangle = \cdots = \langle \hat{j}_N \rangle = \langle \hat{j}_r \rangle \equiv \langle \hat{j} \rangle.  
\end{align}

\subsection{Scaling theory}
\label{sec:scaling}

The behavior of $\langle \hat{j} \rangle$ changes depending on the transport regime of the system, and can be analyzed using scaling theory. From basic microscopic transport theory, the variance of a local inhomogeneity $\langle \Delta x^2 \rangle$ grows in space as a function of time $t$ as
\begin{align}
\label{eq:variance}
\langle \Delta x^2 \rangle = 2D\,t^{2\delta}\,,
\end{align} 
where $\delta$ ($0 < \delta \leq 1$) is the transport coefficient, and $D$ as the diffusion coefficient. The value of $\delta$ is set by how perturbations propagate across the system. This parameter can also be extracted by studying the scaling of the expectation value of the current in the NESS (from here on, unless otherwise specified, all expectation values are taken in the NESS) as a function of chain size as 
\begin{align}
\label{eq:scaling}
\langle \hat{j} \rangle \propto \frac{1}{N^{\nu}}
\end{align}
where $\nu \geq 0$ is the transport exponent. The parameters $\delta$ and $\nu$ are related by $\delta = 1 / (1 + \nu)$ \cite{LiScaling2003}. 

Different transport regimes are identified based on the value of $\nu$ as follows: $\nu = 0$ implies no dependence on system size and occurs when excitations in the system propagate without scattering, i.e., the system behaves as a perfect conductor and transport is {\em ballistic} (also known as {\em coherent}). This regime is expected for integrable systems~\cite{Ilievski2013}. A known exception is the XXZ model for $\Delta \geq 1$, which is integrable yet exhibits nonballistic spin transport \cite{ZnidaricXXZspintransport}. $\nu = 1$ implies a regular {\em diffusive} regime and spin transport in the system obeys Fick's law, so the current across the system is proportional to the gradient of the driving field. The cases $0 < \nu < 1$ and $\nu > 1$ are referred to as {\em anomalous diffusion}, specifically, superdiffusion and subdiffusion, respectively. In these cases, the constant of proportionality (the diffusion coefficient $D$) in Eq.~\eqref{eq:scaling} picks up a dependence on the system size given by $D \propto N^{1 - \nu}$~\cite{Ilievski2013}.

In this work, we use finite-size scaling of the expectation values of the current in the NESS to probe the effect of integrability breaking in Eqs.~\eqref{eq:h_si} and~\eqref{eq:h_sf}. Next, we describe the numerical methods used in our calculations.

\subsection{Solution to the NESS}
\label{sec:solution}

Mathematical properties of the NESS can be obtained from properties of the Liouville superoperator. In order to visualize them, it is convenient to use a vectorization procedure on the density matrix~\cite{LandiFluxRectificationXXZ, EduLinearW}. The procedure consists in concatenating the columns of the density matrix onto a vector. This allows to factorize a Liouville superoperator in matrix form that acts on a vector form of the density matrix. Using a matrix representation of the superoperator allows the Lindblad master equation to be written as
\begin{align}
\label{eq:linmaster}
\frac{d|\hat{\rho}\rangle\rangle}{dt} = \hat{W}|\hat{\rho}\rangle\rangle,
\end{align}
where $|\cdot\rangle\rangle$ is a vectorized matrix built by concatenating its columns, and $\hat{W}$ is the matrix representation of the superoperator in Eq.~\eqref{eq:supop}. The master equation [Eq.~\eqref{eq:lme}] can be expressed in such a way because the vectorization procedure is a linear operation, and  all the terms in Eq.~\eqref{eq:lind1} are of the form $\hat{A}\hat{B}\hat{C}$, where $\hat{A}$, $\hat{B}$, and $\hat{C}$ are matrices. In light of this, the following relation can be used to obtain Eq.~\eqref{eq:linmaster} \cite{LandiFluxRectificationXXZ}:
\begin{align}
\label{eq:vectorized}
|\hat{A}\hat{B}\hat{C}\rangle\rangle = (\hat{C}^{T} \otimes \hat{A})|\hat{B}\rangle\rangle.
\end{align}
From Eq.~\eqref{eq:lind1}, this relation is the only one needed to reduce the Lindblad master equation to Eq.~\eqref{eq:linmaster} in terms of the density matrix and the Pauli spin matrices. 
We use two methods to solve for the NESS. In the first one, we solve a system of linear equations using a matrix representation of the superoperator $\mathcal{W}$ from Eq.~\eqref{eq:supop}, limited only by the accessible system sizes; while the second one is based on time-dependent Matrix Product States (tMPS) \cite{Schollwock2011, Verstraete2008} in combination with a fourth-order Suzuki-Trotter decomposition of the Liouville propagator. We provide a brief description of both methods in Appendices~\ref{sec:directmethod} and~\ref{sec:mpos}.

\section{Numerical results for the NESS}
\label{sec:results}

Here we report the results obtained within the open system's framework reviewed in the previous section, using the methods described in Appendices~\ref{sec:directmethod} and~\ref{sec:mpos}. We focus on the expectation values of the magnetization, and on the expectation values of the current operator as functions of system size (to extract the transport exponent), in both the $\hat{H}_{\textrm{SI}}$ and $\hat{H}_{\textrm{SF}}$ models. 

\subsection{Transport in the single-impurity model $\mathbf{\hat{H}}_{\textrm{SI}}$}
\label{sec:impurity}

In Sec.~\ref{sec:lss}, we showed that a sufficiently strong (but finite) impurity field is able to break the integrability of the XXZ model as seen from the behavior of the level spacing statistics, regardless of the $\mathcal{O}(1)$ nature of the perturbation~\cite{Santos:2004, santos2011domain, torres2014local, XotosIncoherentSIXXZ}. In this section, we investigate the transport of spin excitations in this setup. 

\begin{figure}[!b]
\fontsize{13}{10}\selectfont 
\centering
\includegraphics[width=\columnwidth]{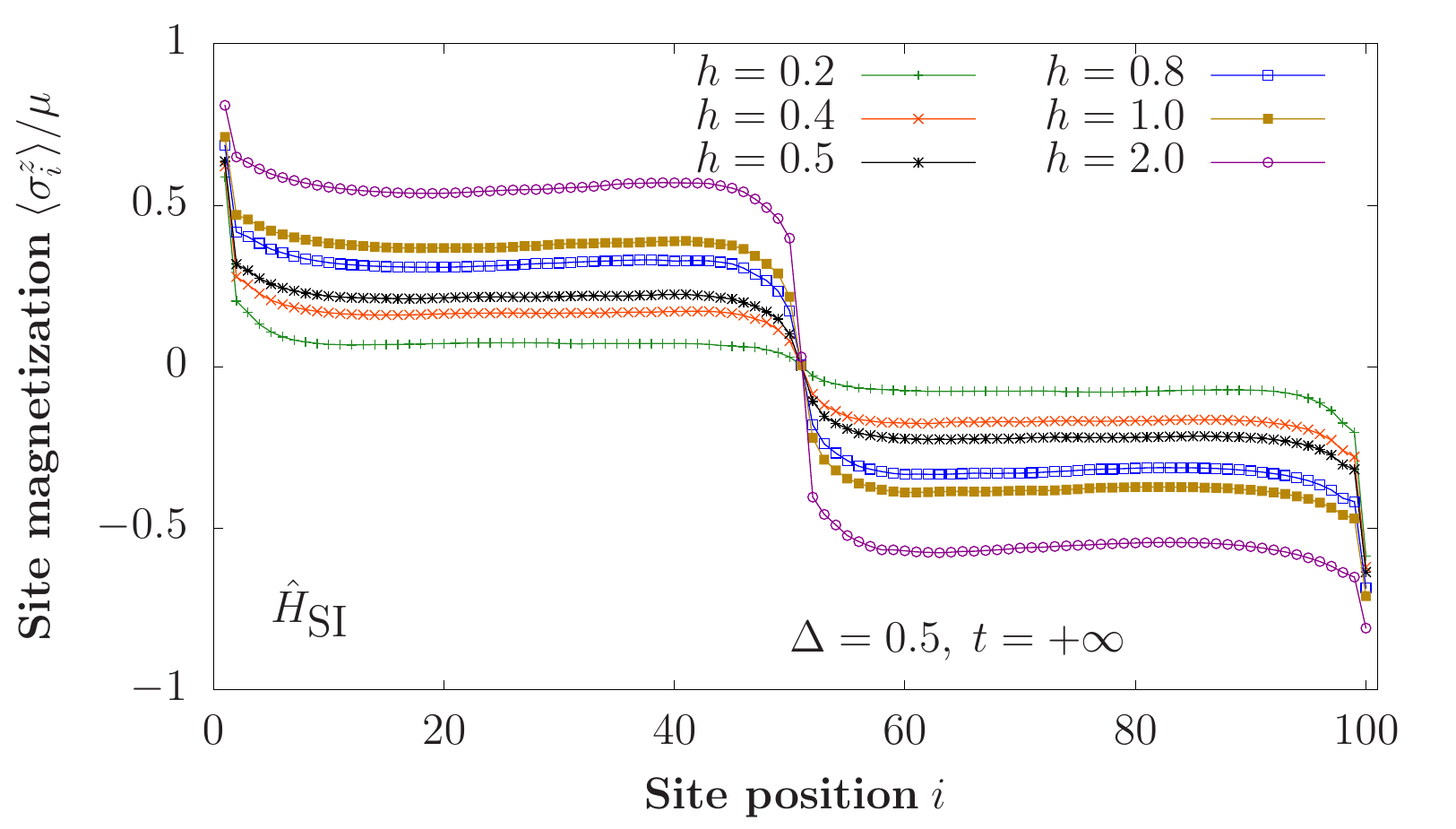}
\caption{Magnetization profile in the nonequilibrium steady state of the anisotropic Heisenberg model in the presence of a single magnetic impurity with different values of $h$ [see Eq.~\eqref{eq:h_si}]. The profiles were obtained for chains with $N = 100$, $\Delta = 0.5$, $\gamma = 1.0$, and $\mu = 0.005$.}
\label{fig:6}
\end{figure}

It is enlightening to first look at the magnetization profile across the chain in the NESS in the presence of the impurity perturbation. In fact, the profile itself is determined by the transport regime of the system. In Fig.~\ref{fig:6}, we show the expectation value of $\hat{\sigma}_{z}$ in the NESS, as a function of site positions, for different values of the impurity strength $h$. The profiles reveal strong boundary effects induced by the driving at the edges of the chain, and are nearly {\em flat} in the bulk of the chain, with the exception of the site where the impurity is located. The ``kink'' at the latter point is larger the stronger the impurity field. The {\em flat} profiles in Fig.~\ref{fig:6} are a first indicator that transport is ballistic, as seen in integrable models such as the unperturbed XXZ chain~\cite{MarkoStaggered1}.

\begin{figure}[!t]
\fontsize{13}{10}\selectfont 
\centering
\includegraphics[width=\columnwidth]{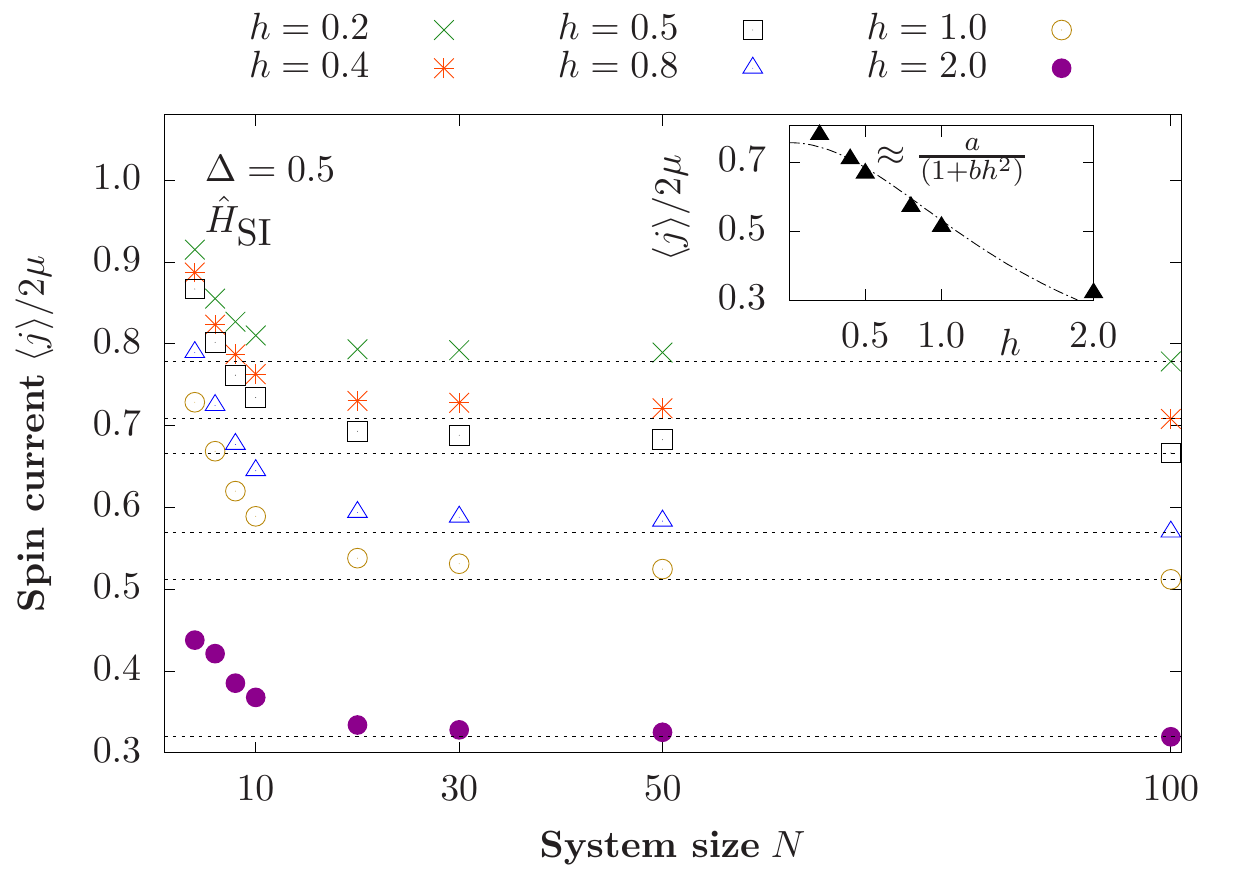}
\caption{Scaling of the expectation value of the current operator in the nonequilibrium steady state of the anisotropic Heisenberg model in the presence of a single magnetic impurity [see Eq.~\eqref{eq:h_si}], plotted as a function of system size ($N=4,\cdots,100$), for $\Delta=0.5$ and different values of $h$. The driving parameters are $\gamma = 1.0$ and $\mu = 0.005$.}
\label{fig:7}
\end{figure}

Next, we quantify how the current in the NESS scales with increasing system size. In Fig.~\ref{fig:7}, we plot $\langle \hat{j} \rangle$ vs $N$ for $\Delta = 0.5$ and different values of $h$. Transport in the XXZ model is ballistic for any $0 < \Delta < 1$, a regime that is expected to change to incoherent, either diffusive or anomalous, when integrability is broken. We chose $\Delta=0.5$ because the system is in the strongly-interacting regime, and obtaining the NESS numerically is not as difficult as for $\Delta \approx 1$. The main observation in Fig.~\ref{fig:7} is that, for sufficiently large system sizes, $\langle \hat{j} \rangle$ becomes independent of $N$, a property of systems that exhibit coherent/ballistic transport. 

Our high-temperature nonequilibrium calculations indicate that, even though a single magnetic impurity breaks the integrability of the XXZ chain as seen from the probability distribution of energy level spacings (Fig.~\ref{fig:lss}), transport remains ballistic and the system behaves as a perfect conductor. This becomes apparent in the scaling of the spin current only for sufficiently large system sizes, see Fig.~\ref{fig:7}, in analogy with the integrable XXZ case \cite{ZnidaricXXZspintransport}. We stress that this behavior persists for all the values of $h$ studied, and that we expect it to persists for any finite nonvanishing magnetic impurity strength (for $h=0$ one has an integrable XXZ chain, and for $h=\infty$ one has two disconnected integrable XXZ chains). This is the first example known to us in which a quantum many-body system exhibits a Wigner-Dyson level spacing distribution and displays coherent transport. The latter can be understood to be the result of excitations traveling in a ballistic fashion on either side of the integrability breaking defect and scattering only at the impurity site. 

The inset in Fig.~\ref{fig:7} shows the scaling of the steady-state spin current, for $N = 100$, with the impurity strength. $\langle \hat{j} \rangle$ vs $h$ can be well fitted with the function $a / (1 + bh^2)$, an ansatz that follows from results for the noninteracting case discussed in Appendix~\ref{ap:nonint}. The main effect of increasing the magnitude of $h$ is to decrease the magnitude of  $\langle \hat{j} \rangle$, while transport remains ballistic.

The results reported here suggest that a single impurity is not sufficient to render transport incoherent, despite the fact that it is enough to render the system quantum chaotic, as indicated by the distribution of energy levels. In Sec.~\ref{sec:staggered}, we revisit spin transport in the XXZ model in the presence of a staggered field, to contrast the results with those obtained in this section.

\subsection{Transport in the staggered-field model $\mathbf{\hat{H}}_{\textrm{SF}}$}
\label{sec:staggered}

While it is known that the gapless XXZ model ($0 < \Delta < 1$) exhibits ballistic spin transport, and it is therefore an ideal conductor~\cite{ZnidaricXXZspintransport, juan2014a}, breaking integrability by means of a staggered magnetic field renders the system chaotic and spin transport  becomes diffusive~\cite{MarkoStaggered1}. We revisit transport in the $\hat{H}_{\textrm{SF}}$ model [see Eq.~\eqref{eq:h_sf}] to contrast it with that in the $\hat{H}_{\textrm{SI}}$ model [see Eq.~\eqref{eq:h_si}].

\begin{figure}[!b]
\centering
\includegraphics[width=\columnwidth]{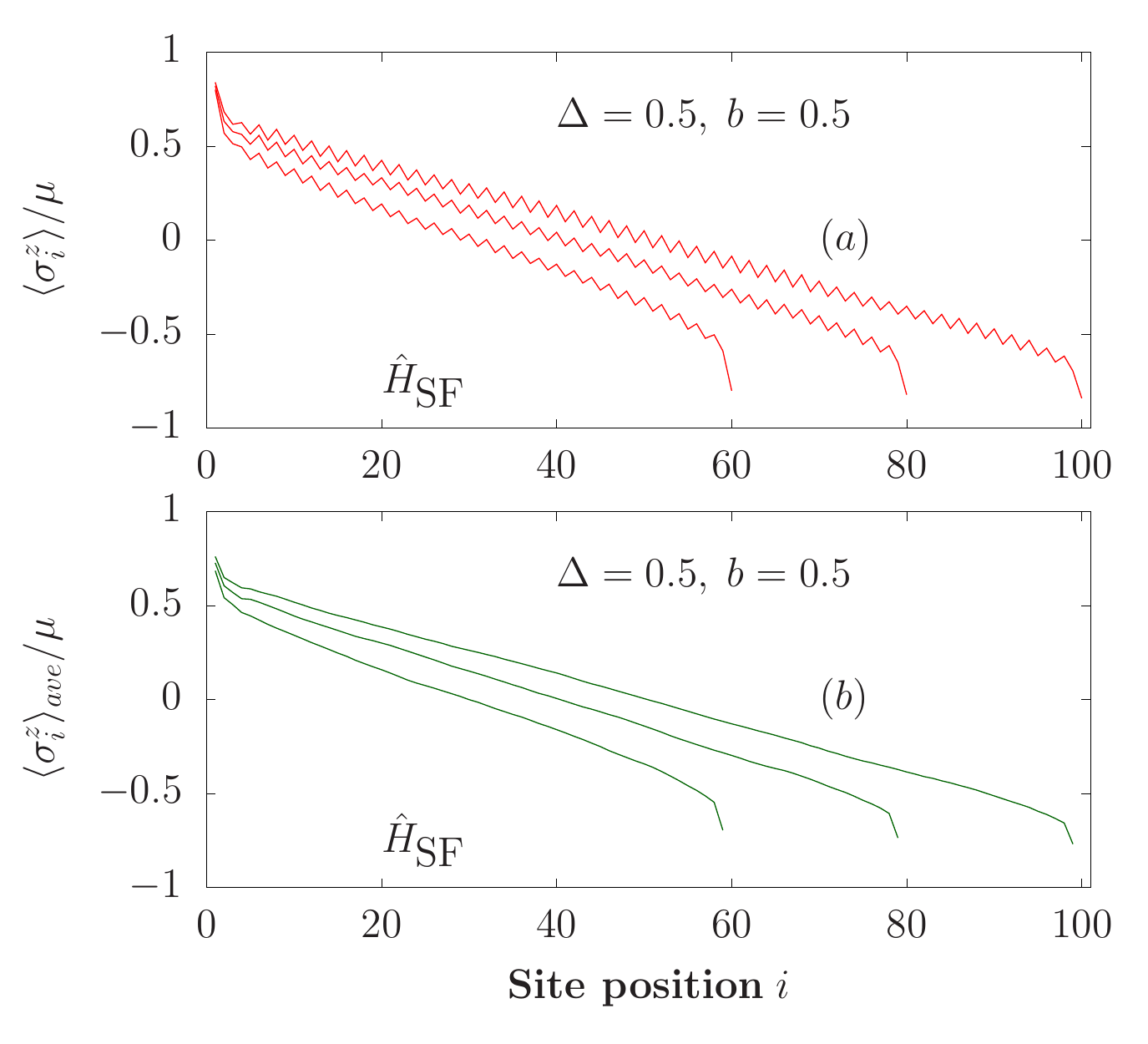}
\caption{Magnetization profile of the nonequilibrium steady state of the anisotropic Heisenberg model in the presence of a staggered magnetic field with $b = 0.5$. The results were obtained for $\Delta = 0.5$, $\gamma = 1.0$, and $\mu = 0.001$. (a) Magnetization, and (b) average magnetization [see Eq.~\eqref{eq:avemag}].}
\label{fig:8}
\end{figure}

Figure~\ref{fig:8}(a) shows the magnetization profile in the NESS of the $\hat{H}_{\textrm{SF}}$ model for $\Delta=0.5$, $b=0.5$, and different chain sizes. Unlike the magnetization profile in the NESS for the $\hat{H}_{\textrm{SI}}$ model, the staggered field induces a ramplike linear profile in the magnetization across the chain. The small oscillations of the magnetization are due to the presence of the staggered field. In Fig.~\ref{fig:8}(b), we show the average
\begin{align}
\label{eq:avemag}
\langle \hat{\sigma}^z_i \rangle_{ave} = \left(\langle \hat{\sigma}^z_i \rangle + \langle \hat{\sigma}^z_{i+1} \rangle\right) / 2.
\end{align}
Figure~\ref{fig:8}(b) makes apparent that, aside from boundary effects, the magnetization profile is linear.

\begin{figure}[!t]
\centering
\includegraphics[width=\columnwidth]{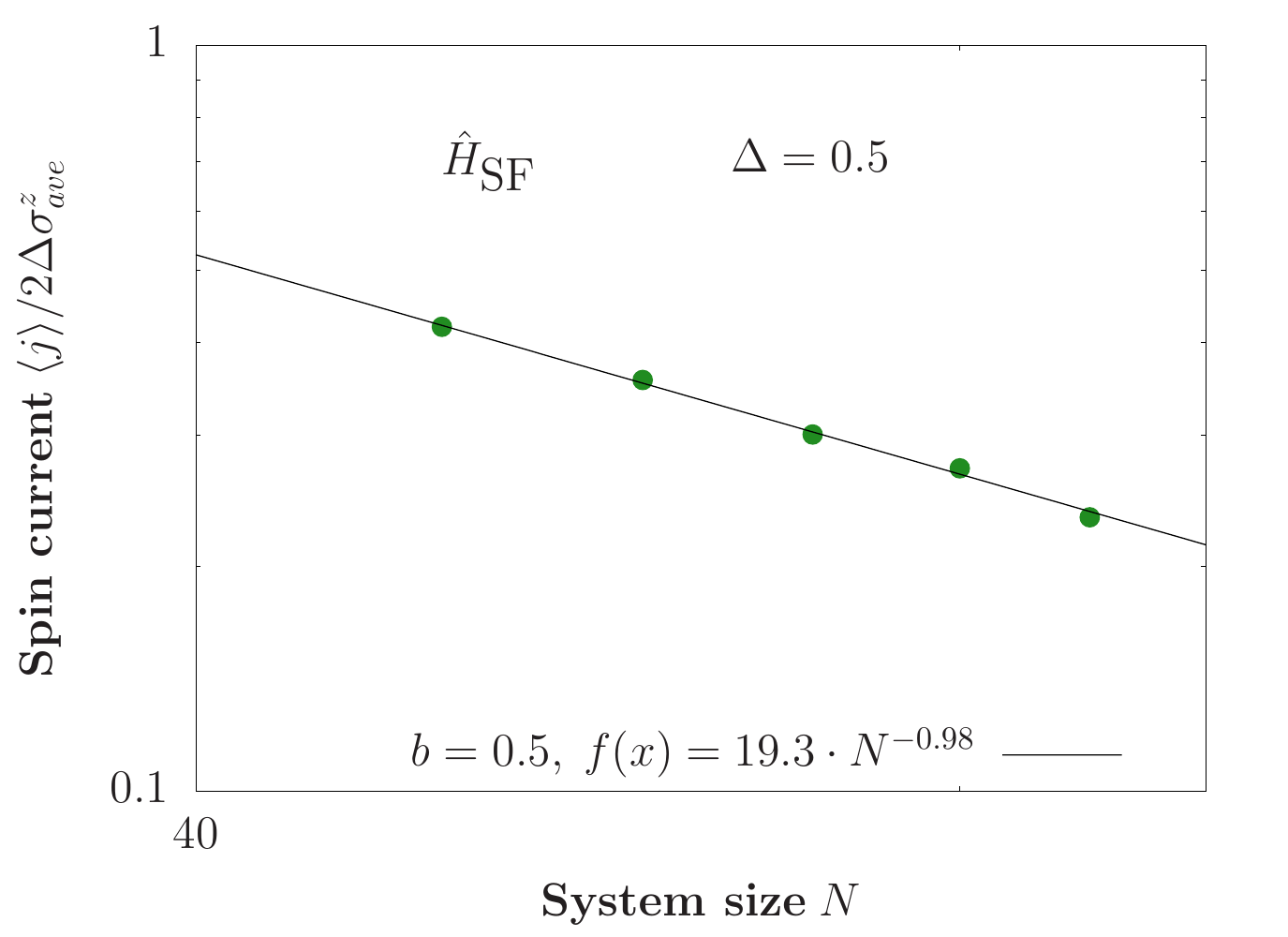}
\caption{Scaling of the spin current in the NESS of the staggered-field model as a function of system size ($N=60,70,80,90,100$), for $\Delta=0.5$ and $b = 0.5$ (same parameters as in Fig.~~\ref{fig:8}). The driving parameters are $\gamma = 1.0$ and $\mu = 0.001$. To reduce finite-size effects, in our calculations we discard the five leftmost and the five rightmost sites of the chains.}
\label{fig:9}
\end{figure}

Figure~\ref{fig:9} shows results for the finite-size scaling of the spin current in the NESS of the $\hat{H}_{\textrm{SI}}$ model, for the same parameters used in Fig.~\ref{fig:8}. We obtain the diffusion parameters, $D = 19.3$ (the diffusion coefficient) and $\nu = 0.98$, from
\begin{align}
\label{scaling_sf}
\frac{\langle \hat{j} \rangle}{2\Delta \hat{\sigma}^z_{ave}} = \frac{D}{(N-10)^{\nu}}.
\end{align}
Our results show that the current obeys the diffusion equation (Fick's law). They are in agreement with the results in Ref.~\cite{MarkoStaggered1}. 

To check that we are working in the linear response regime, for both the $\hat{H}_{\textrm{SI}}$ and $\hat{H}_{\textrm{SF}}$ models, we studied the spin current as a function of the driving strength. As discussed in Appendix~\ref{ap:error}, in the regime of small values of $\mu$, the current in our calculations depends linearly on the driving parameter. 

\section{Kubo linear response theory}
\label{sec:linearresponse}

In the previous section, we studied two nonintegrable models [described by the Hamiltonians in Eqs.~\eqref{eq:h_si} and~\eqref{eq:h_sf}] displaying contrasting transport properties. Specifically, the single impurity model (with $\hat{H}_{\textrm{SI}}$) displays coherent transport, while the staggered field model (with $\hat{H}_{\textrm{SF}}$) displays diffusive transport, despite the fact that both exhibit quantum chaotic energy spacing distributions as predicted by random matrix theory. To understand whether the differences found are {\em real} and not, e.g., an artifact of the microscopic details of the $\mathcal{L}_{l,r}$ dissipators in Eq.~\eqref{eq:lind2}, we turn to Kubo linear response theory for closed quantum systems.

Within linear response theory, the real part of the conductivity can be written as ($\hbar = 1$ and $k_B = 1$) \cite{kubo1957statistical, kubo1957statistical2, ShastryKubo2008, RigolShastry2008}
\begin{align}
\label{eq:kuboformula}
\textrm{Re}[\sigma_N(\omega)] &= \pi D_N\delta(\omega) + \nonumber \\
\frac{\pi}{N}&\left(\frac{1-e^{-\beta \omega}}{\omega}\right)\sum_{\epsilon_n \neq \epsilon_m} p_n|J_{nm}|^2\delta(\epsilon_m - \epsilon_n - \omega),
\end{align}
where $D_N$ is known as the Drude weight or spin stiffness, $\beta$ is the inverse temperature, $p_n= e^{-\beta\epsilon_n} / Z$ is the Boltzmann weight of eigenstate $\ket{n}$ with energy $\epsilon_n$, and $Z$ is the partition function. $J_{nm}$ are the matrix elements of the {\em total} spin current operator in the energy eigenbasis
\begin{align}
\label{eq:totalcurrent}
\hat{J} = \sum_i \hat{j}_i,
\end{align}
with the sum adjusted properly depending on whether the system has periodic or open boundary conditions. Here, $\hat{j}_i$ is the local spin current operator from Eq.~\eqref{eq:spincurrent}. 

The Drude weight can be calculated using the expression
\begin{align}
\label{eq:drude1}
D_N = \frac{1}{N}\left[\langle -\hat{\Gamma} \rangle - \sum_{\epsilon_n \neq \epsilon_m} \frac{p_n - p_m}{\epsilon_m - \epsilon_n}|J_{nm}|^2\right],
\end{align}
where $\hat{\Gamma}$ is the so-called stress tensor operator~\cite{ShastrySum2006}, which is identical to the kinetic energy operator $\hat{T} = \sum_i \alpha\,\left( \hat{\sigma}^x_{i}\hat{\sigma}^x_{i+1} + \hat{\sigma}^y_{i}\hat{\sigma}^y_{i+1} \right)$ in the models we consider. In one dimension and for sufficiently high temperatures (in the absence of superconductivity \cite{ShastryKubo2008, RigolShastry2008}), the Drude weight can also be obtained using the expression~\cite{Heidrich-Meisner03}
\begin{align}
\label{eq:drude2}
\bar{D}_N = \frac{\beta}{N}\sum_{\epsilon_n = \epsilon_m} p_n|J_{nm}|^2.
\end{align}

In the thermodynamic limit, Eq.~\eqref{eq:kuboformula} leads to the decomposition $\textrm{Re}[\sigma_{\infty}(\omega)] = \pi D_{\infty}\delta(\omega) + \sigma_{reg}(\omega)$, where $D_{\infty} = \lim_{N \rightarrow \infty}D_N = \lim_{N \rightarrow \infty}\bar{D}_N$ and $\sigma_{reg}(\omega)$ is the regular part of the conductivity. A nonzero $D_{\infty}$ signals that transport is ballistic, the current-current correlation function does not vanish in the limit of infinite time. This is a property of integrable systems. In systems that display diffusive transport, expected for nonintegrable systems, $D_{\infty}=0$. 

Equations~\eqref{eq:kuboformula} through~\eqref{eq:drude2} are usually evaluated in systems with translation invariance. In systems with open boundary conditions, such as the ones for which the NESS was evaluated in the previous section, obtaining $D_{\infty}$ is subtle. In such systems, the position operator
\begin{align}
\label{eq:position}
\hat{X} \coloneqq \sum_k k\,\hat{\sigma}^{+}_k\hat{\sigma}^{-}_k
\end{align}
is well defined \cite{ShastrySum2006}.  $\hat{X}$ can be used to define the total current operator as $\hat{J} = i[\hat{X}, \hat{H}]$, where $\hat{H}$ is the Hamiltonian; and the stress tensor operator as $\hat{\Gamma} = -i[\hat{X}, \hat{J}]$ \cite{RigolShastry2008}. If one uses these relations to evaluate the matrix elements of the total current operator, one finds that $J_{nm} = i\braket{n | \hat{X}\hat{H} | m} - i\braket{n | \hat{H}\hat{X} | m} = i(\epsilon_m - \epsilon_n)\braket{n | \hat{X} | m}$, which implies that $D_{N}$ and $\bar{D}_N$ are exactly zero~\cite{RigolShastry2008}. This implies that, in systems with open boundary conditions, $\lim_{N \rightarrow \infty}D_N = \lim_{N \rightarrow \infty}\bar{D}_N=0$ irrespective of whether the system is integrable or not, in disagreement with what is known for systems with periodic boundary conditions. Such a disagreement may lead one to question whether the Drude weight obtained from this picture [Eqs.~\eqref{eq:kuboformula}--\eqref{eq:drude2}] is a meaningful thermodynamic quantity. The fact that it is was argued for in Ref.~\cite{RigolShastry2008}.

A central finding of Ref.~\cite{RigolShastry2008} is that, in order to obtain $D_{\infty}\neq0$ in integrable systems with open boundary conditions and conciliate the result with the one obtained in systems with periodic boundary conditions, one needs to study the behavior of the finite frequency part of the Kubo formula [the second term in Eq.~\eqref{eq:kuboformula}]. In the thermodynamic limit, a peak develops at zero frequency from the collapse of peaks located at finite (size dependent) frequencies in finite-size systems.

Even in the presence of periodic boundary conditions, one can see that a similar analysis is needed for $\hat{H}_{\textrm{SI}}$. Having an impurity with a very strong field ($h\rightarrow\infty$) is equivalent to having open boundary conditions. Also, in the noninteracting limit ($\Delta=0$) for which transport must be ballistic, the presence of the impurity breaks the $k,-k$ degeneracy in the single-particle spectrum resulting in $D_N = \bar{D}_N=0$. The latter remains true for $\Delta\neq0$. Next, we study the finite-frequency part of Eq.~\eqref{eq:kuboformula} in the single impurity model at high temperature.

\subsection{Numerical Results}
\label{subsec:linearresponse}

We compute the finite-frequency part of Eq.~\eqref{eq:kuboformula} within the grand-canonical ensemble (at zero chemical potential), for which finite-size effects are expected to be the smallest in the presence of translational invariance~\cite{iyer_srednicki_15}. We only study chains with an even number of lattice sites given the known presence of strong even-odd effects at high temperature~\cite{BrenigTypicality2015}.  Since we are interested in the high temperature regime (we take $\beta = 0.001$ in all our calculations), the calculation requires the evaluation of all the eigenenergies and eigenvectors of the Hamiltonian. This is achieved using full exact diagonalization, for which the accessible system sizes with our computational resources are $N\lesssim18$.

\begin{figure}
\fontsize{13}{10}\selectfont 
\centering
\includegraphics[width=0.85\columnwidth]{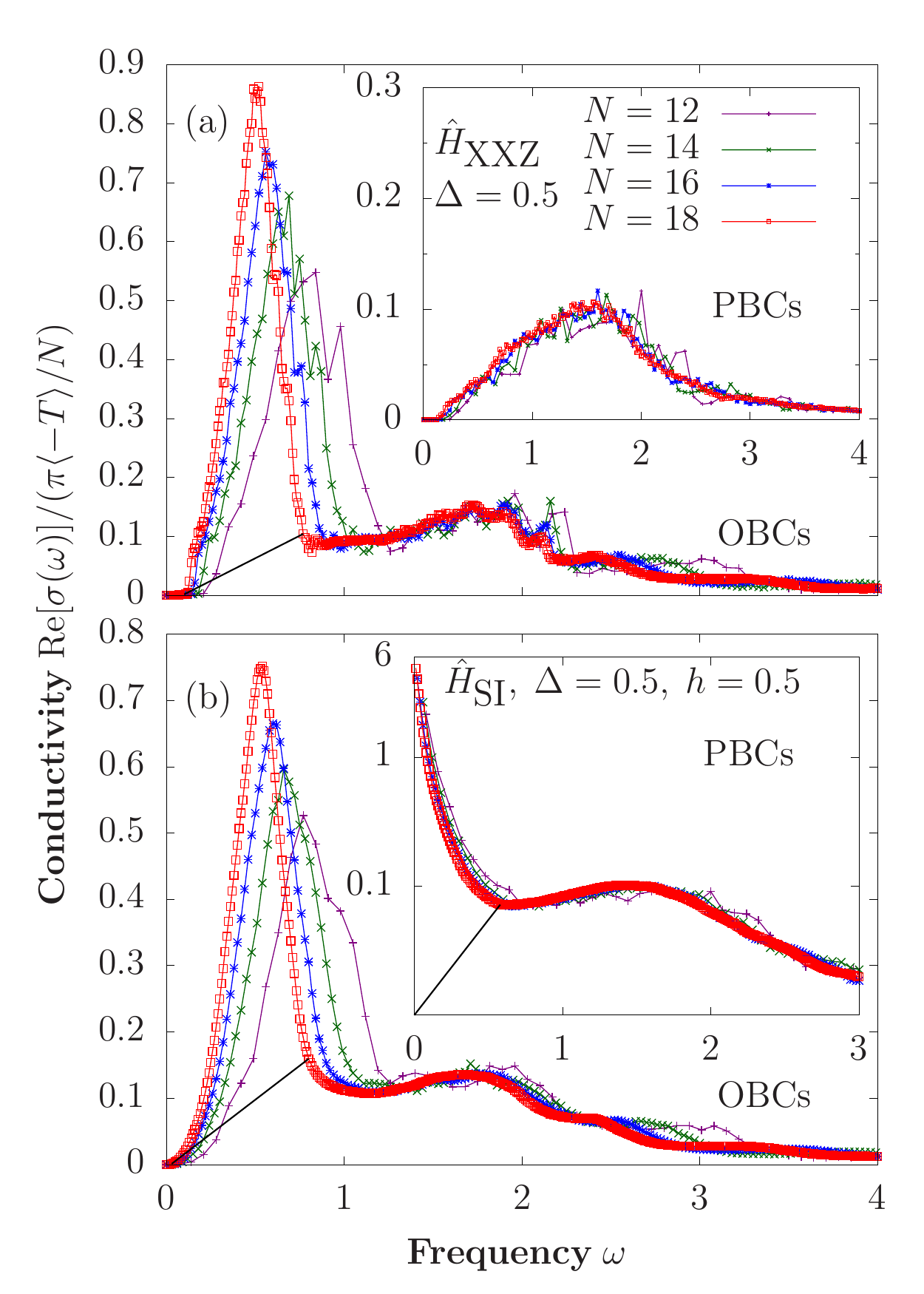}
\caption{Finite-frequency part of the spin conductivity [the second term in Eq.~\eqref{eq:kuboformula}]. (a) Integrable $\hat{H}_{\textrm{XXZ}}$ model in the gapless phase, $\Delta = 0.5$, in chains with (main panel) open boundary conditions and (inset) periodic boundary conditions. (b) Single impurity model $\hat{H}_{\textrm{SI}}$, for $\Delta = 0.5$ and $h=0.5$, in chains with (main panel) open boundary conditions and (inset) periodic boundary conditions (linear-log scale). The results were obtained at very high temperature $\beta = 0.001$. The straight lines in the main panels and in the inset in (b), shown only for $N=18$, are approximate delimiters for the bottom of the large low-frequency peak as suggested by the smooth curves in the inset in (a).}
\label{fig:11}
\end{figure}

In Fig.~\ref{fig:11}(a) and its inset, we show the finite-frequency part of the conductivity for XXZ chains with open and periodic boundary conditions, respectively. A binning procedure was used in order to obtain smooth curves. The size of the frequency bins is selected to be large enough so that the bins contain a large enough number of the discrete frequencies of the system, but small enough so that the results are robust against changes of the bin size. In our simulations, we used bin sizes of 0.001-0.1 depending on the dimension of the Hilbert space for each magnetization subsector. The curves are normalized to satisfy the sum rule
\begin{align}
\label{eq:sumrule}
\int^{\infty}_0\textrm{Re}[\sigma(\omega)]d\omega &= \frac{\pi \langle -\hat{T} \rangle}{2N} \nonumber \\
\implies &\frac{N}{\pi \langle -\hat{T} \rangle} \int^{\infty}_0\textrm{Re}[\sigma(\omega)]d\omega = \frac{1}{2},
\end{align}
\noindent
so that the area under the curves is $1 / 2$.

\begin{figure}
\fontsize{13}{10}\selectfont 
\centering
\includegraphics[width=0.85\columnwidth]{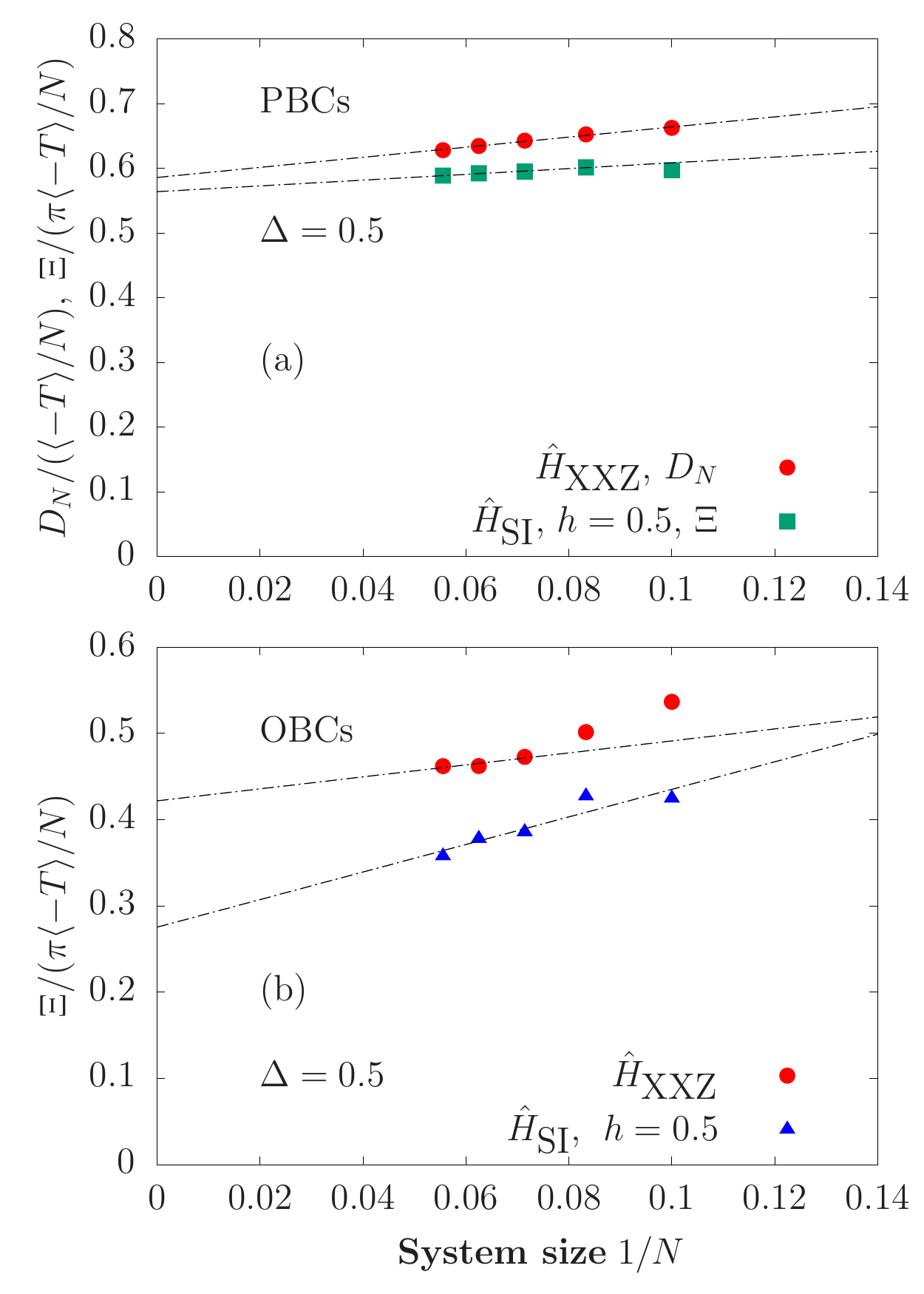}
\caption{Finite-size scaling analysis (up to $N = 18$) of (a) the Drude weight for the $\hat{H}_{\textrm{XXZ}}$ model and the weight of the lowest frequency peak $\Xi$ for the $\hat{H}_{\textrm{SI}}$ model in chains with periodic boundary conditions, and (b) the weight of the lowest frequency peak $\Xi$ for the $\hat{H}_{\textrm{XXZ}}$ and $\hat{H}_{\textrm{SI}}$ models in chains with open boundary conditions. All results were obtained at very high temperature $\beta = 0.001$.}
\label{fig:12}
\end{figure}

The main panel and the inset in Fig.~\ref{fig:11}(a) show that there is a stark contrast between the finite-frequency part of $\textrm{Re}[\sigma_N(\omega)]$ in the integrable XXZ model depending of whether the chains have open or periodic boundary conditions (see also Fig.~1 in Ref.~\cite{RigolShastry2008}). For periodic boundary conditions, the finite-frequency part exhibits a smooth behavior that is nearly size-independent. The Drude weight in that case, shown in Fig~\ref{fig:12}(a), extrapolates to a nonzero value in the thermodynamic limit. 

For open boundary conditions, a large sharp peak can be seen at low frequencies (smaller sharp peaks occur at higher frequencies) on top of an otherwise smooth part that resembles that of the system with periodic boundary conditions. This sharp peak moves toward smaller frequencies with increasing system size ($\omega_\text{peak}\propto 1/N$, see Appendix~\ref{ap:nonint} and Ref.~\cite{RigolShastry2008}), so one expects it to be at zero frequency in the thermodynamic limit. The area under this peak, and above the smooth curve seen in the system with periodic boundary conditions, extrapolates to a finite value in the thermodynamic limit. The latter is shown in Fig~\ref{fig:12}(b), where $\Xi$ is two times the area under the peak and above of the straight line in Fig.~\ref{fig:11}(a). The extrapolated value obtained for $\Xi$ in the thermodynamic limit is smaller than the one obtained for $D_\infty$ in systems with periodic boundary conditions in Fig~\ref{fig:12}(a). The expectation for systems with open boundary conditions is that other peaks at higher frequencies, which are also $\propto 1/N$, will collapse to $\omega=0$ in the thermodynamic limit, and their added weight will be identical to the Drude weight obtained in systems with periodic boundary conditions~(see Appendix~\ref{ap:nonint} and \cite{RigolShastry2008}). This is how a nonvanishing Drude weight appears in systems with open boundary conditions, for which $D_N=\bar D_N=0$ for any $N$.

In the main panel in Fig~\ref{fig:11}(b), we show the finite-frequency part of $\textrm{Re}[\sigma_N(\omega)]$ in the single-impurity model for chains with open boundary conditions. The curves are very similar to those obtained for the integrable XXZ model in Fig~\ref{fig:11}(a). Also, the extrapolation shown in Fig~\ref{fig:12}(b) suggests that the area under the large low-frequency peak is finite in the thermodynamic limit as for the integrable XXZ model. The inset in Fig~\ref{fig:11}(b) shows the results for the finite-frequency part of $\textrm{Re}[\sigma_N(\omega)]$ in the $\hat{H}_{\textrm{SI}}$ model for chains with periodic boundary conditions. They are in stark contrast to those for the XXZ chain in systems with periodic boundary conditions, and have features present in the results for chains with open boundary conditions. A smooth, nearly system-size independent, part is seen at frequencies $\omega>0.5$, and a sharp peak is seen about $\omega=0$. The width of the sharp peak decreases with increasing system size, while its area extrapolates to a finite value in the thermodynamic limit. In Fig.~\ref{fig:12}(a) we show the extrapolation of $\Xi$, which gives a result in the thermodynamic limit that is very close to the Drude weight obtained in systems with periodic boundary conditions in the absence of the impurity. This suggests that, in the thermodynamic limit, the low-frequency peak collapses to $\omega=0$ resulting in a nonzero Drude weight. Our results for the $\hat{H}_{\textrm{SI}}$ model, both in systems with open and periodic boundary conditions, indicate that transport in the $\hat{H}_{\textrm{SI}}$ model is coherent, in agreement with our boundary-driven calculations from Sec.~\ref{sec:impurity}. 

We should mention that there is an earlier study of the finite-frequency part of $\textrm{Re}[\sigma_N(\omega)]$ in the $\hat{H}_{\textrm{SI}}$ model for chains with periodic boundary conditions~\cite{XotosIncoherentSIXXZ}. The results reported in that work are similar to those reported in the inset in Fig~\ref{fig:11}(b). However, the low-frequency peak whose width vanishes with increasing system size was interpreted as indicating incoherent transport with a relaxation time $\tau\propto N$.  Similar results and conclusions to those in Ref.~\cite{XotosIncoherentSIXXZ} were reported in Refs.~\cite{Metavitsiadis2010, Metavitsiadis2011} for energy transport in the presence of an impurity.
 
\section{Summary and outlook}
\label{sec:conclusions}

Integrability is known to be fragile against perturbations. It is still remarkable that a single impurity can break integrability in an $N\rightarrow\infty$ chain~\cite{Santos:2004, santos2011domain, torres2014local, XotosIncoherentSIXXZ}. This can be understood in view of the fact that an $O(1)$ local integrability breaking perturbation can mix exponentially many extended eigenstates of an integrable model and produce a Wigner-Dyson level spacing distribution typical of quantum chaotic models. Since the quantum chaotic models studied to date exhibit incoherent transport, a Wigner-Dyson level spacing distribution is usually assumed to mean incoherent transport. 

In this work we have studied a model, the first one known to us, for which this intuition does not apply. We showed that, while a single impurity in the XXZ model changes the level spacing distribution from Poisson to Wigner-Dyson, it does not change the nature of spin transport in the chain from coherent (for $0<\Delta<1$) to incoherent. We discussed this both in the context of transport in nonequilibrium steady states and in the context of Kubo linear response theory. We argued that this conclusion applies to chains with open and periodic boundary conditions. Our results hint that the equilibration properties of the single impurity model should be anomalous. The fact that models with single impurities can display anomalies in their approach to equilibrium is a topic that has started to be explored~\cite{fagottiImpurity17, bastianello18}.

It would be interesting to understand the onset of diffusion for systems in which integrability is broken not by a single impurity but by an increasing number of impurities that, e.g., interpolate between the single impurity model and the staggered field model also considered here. The latter was shown to exhibit the expected incoherent transport for a quantum chaotic model. Another interesting question is what happens as one adds impurities in a sequence in which they occupy the central site in empty sections of the chain. These are questions we are currently exploring.

\section{Acknowledgments}
\label{sec:acknow}

We are grateful to S.~R.~Clark, J.~Gemmer, P.~Prelov\v{s}ek, L.~F.~Santos, A.~Scardicchio, B.~S.~Shastry, A.~Silva, V.~K. Varma, and M.~\v{Z}nidari\v{c} for fruitful discussions. M.B. and J.G. acknowledge the DJEI/DES/SFI/HEA Irish Centre for High-End Computing (ICHEC) for the provision of computational facilities and support, project TCPHY104B, and the Trinity Centre for High Performance Computing. This work was supported by an SFI-Royal Society University Research Fellowship (J.G.), the Royal Society (M.B.), and the National Science Foundation Grant No.~PHY-1707482 (M.R.). This project received funding from the European Research Council (ERC) under the European Union's Horizon 2020 research and innovation program (grant agreement No.~758403), and was supported in part by the National Science Foundation under Grant No.~PHY-1748958 (J.G. and M.R.). E. M. was supported by the EPSRC Programme Grant DesOEQ (EP/P009565/1).

\appendix

\section{Numerical evaluation of nonequilibrium steady states}
\label{sec:numerical}

\subsection{Exact numerical approach to the solution of the nonequilibrium steady state}
\label{sec:directmethod}

Using the vectorized form of the density matrix described in Sec.~\ref{sec:solution}, one can write a matrix representation of the Liouville superoperator, and combine operations of the form in Eq.~\eqref{eq:vectorized} in order to factorize this operator from the density matrix. In this picture, Eq.~\eqref{eq:supop} transforms to 
\begin{align}
\label{eq:vecformsupop}
\hat{W}|\hat{\rho}_{\textrm{NESS}}\rangle\rangle = 0,
\end{align}
where $\hat{W}$ is a non-Hermitian matrix of dimension $d_{\mathcal{H}}^2$ and $|\hat{\rho}_{\textrm{NESS}}\rangle\rangle$ is the vector form of the density matrix representing the NESS, with the same dimension. At this point it is clear that, given that the Hilbert space dimension is effectively increased by a power of 2, the computational cost of studying interacting open quantum systems is immensely higher than in closed quantum systems.

The solution of Eq.~\eqref{eq:vecformsupop} is found by directly solving the system of linear equations constrained to the trace preserving property of the density matrix
\begin{align}
\label{eq:trace1}
\langle\langle\mathds{1}|\hat{\rho}\rangle\rangle = \textrm{Tr}(\hat{\rho}) = 1,
\end{align}
where $|\mathds{1}\rangle\rangle$ is the vectorized identity. 

One can then define~\cite{EduLinearW}
\begin{align}
\label{eq:rescaleW}
\widetilde{W} = \hat{W} + |0\rangle\rangle\langle\langle\mathds{1}|,
\end{align}
such that
\begin{align}
\label{eq:linearsolve}
\widetilde{W}|\hat{\rho}_{\textrm{NESS}}\rangle\rangle &= \hat{W}|\hat{\rho}_{\textrm{NESS}}\rangle\rangle + |0\rangle\rangle\langle\langle\mathds{1}|\hat{\rho}_{\textrm{NESS}}\rangle\rangle, \nonumber \\
\widetilde{W}|\hat{\rho}_{\textrm{NESS}}\rangle\rangle &= |0\rangle\rangle, \nonumber \\
\implies |\hat{\rho}_{\textrm{NESS}}\rangle\rangle &= \widetilde{W}^{-1}|0\rangle\rangle,
\end{align}
where $|0\rangle\rangle$ is the vectorized form of the first state in the Hilbert space. The choice of the matrix $|0\rangle\rangle\langle\langle\mathds{1}|$ is in principle arbitrary, with the only condition that the trace of the density matrix is preserved. In the present case, $|0\rangle\rangle\langle\langle\mathds{1}|$ is a matrix of zeroes, with ones only in the first row in the columns corresponding to the diagonal elements of $\hat{\rho}$.

It is impractical to evaluate $\widetilde{W}^{-1}$ given that, even if $\widetilde{W}$ is sparse, $\widetilde{W}^{-1}$ will not be sparse in general. Therefore, the solution to the linear system is normally tackled by means of direct or indirect methods. In general, direct methods are more expensive in both computational and memory terms. However, indirect methods such as Krylov subspace techniques normally require preconditioning or other additional techniques to attain acceptable numerical convergence with a low number of operations. 

The main drawback of the exact numerical approach is intractability, in light of the $d_{\mathcal{H}}^2$ scaling of the Hilbert space. In our work, we used this method only for small system sizes $N \sim 10$. These system sizes are generally too small to identify the transport regime in boundary driven spin chains. We resort to the tMPS technique, briefly described in Sec.~\ref{sec:mpos}, and use the exact approach to evaluate the numerical fidelity of the results obtained with tMPS.

\subsection{Matrix product states-operators approach to the solution of the nonequilibrium steady state}  
\label{sec:mpos}

In order to solve large system sizes, we use the time-dependent Matrix Product States algorithm to study the evolution of any initial state under Eq.~\eqref{eq:lme}. We start by writing the density matrix of the system in the form
\begin{align}
\label{eq:rhomps}
|\rho \rangle = \sum_{\sigma_1,\cdots,\sigma_N} c_{\sigma_1\cdots\sigma_N}| \sigma_1, \cdots, \sigma_N \rangle,
\end{align}
where there are $d^N$ coefficients $c_{\sigma_1\cdots\sigma_N}$ that describe the state of the system, and $\sigma_{i}$ is the local basis at site $i$. The Pauli basis is a natural and commonly used choice to represent the local basis, such that at site $i$ the local basis is given by
\begin{align}
\label{eq:basis}
\{\sigma_i\} = \left\{ \frac{1}{2}\mathds{1}, \frac{1}{2}\hat{\sigma}^x, \frac{1}{2}\hat{\sigma}^y, \frac{1}{2}\hat{\sigma}^z \right\}.
\end{align}
We use the vectorized form of this local basis, i.e., $\textrm{vec}(\hat{\sigma}^{\nu})$ such that the density matrix operator can be represented as an MPS in the extended Hilbert space. The power of the MPS representation of the density matrix resides on the fact that it provides a sense of locality to the state, while preserving the inherent quantum nonlocality features. To achieve this, Eq.~\eqref{eq:rhomps} has to be expressed in MPS form as
\begin{align}
\label{eq:rhofinalmps}
| \rho \rangle = \sum_{\sigma_1,\cdots,\sigma_N} A^{\sigma_1}A^{\sigma_2} \cdots A^{\sigma_{N-1}}A^{\sigma_N} | \sigma_1,\cdots,\sigma_N \rangle,
\end{align}
where the $A^{\sigma_i}$ are a collection of $d$ matrices of dimension $\chi \times \chi$. This form can be generated from the sequential reshaping and singular value decomposition (SVD) procedures of the $c_{\sigma_1\cdots\sigma_N}$ coefficients \cite{Schollwock2011}.

For the specific case at hand, one can use this representation in a tractable manner and keep the degree of correlations (manifest in the bond dimension $\chi$ of the matrices $A^{\sigma_{i}}$) under control by using an initial product state, say for instance, the identity state; and evolving the system under dynamics that keep the state close to an identity state throughout the evolution as the NESS is reached. From Eq.~\eqref{eq:lind2}, this can be achieved for {\em small} values of $\mu$. Increasing this parameter results in states of the system that are further away from the identity in terms of quantum correlations, i.e., states that require a large bond dimension to be represented with high fidelity; particularly for large system sizes.  

Just like states, operators can be written in MPS form in a representation known as Matrix Product Operators (MPOs). Given that any quantum operator can be expressed as
\begin{align}
\label{eq:mpo1}
\hat{O} = \sum_{\boldsymbol{\sigma},\boldsymbol{\sigma^{\prime}}} c_{(\sigma_1,\cdots,\sigma_N), (\sigma^{\prime}_1,\cdots,\sigma^{\prime}_N)} | \boldsymbol{\sigma} \rangle \langle \boldsymbol{\sigma^{\prime}} |,
\end{align}
with $\boldsymbol{\sigma} \coloneqq | \sigma_1,\cdots,\sigma_N \rangle$, one can decompose $\hat O$ the same way as for an MPS with the double index $\sigma^{}_i \sigma^{\prime}_i$ taking the role of the single index $\sigma_i$ to give
\begin{align}
\label{eq:mpo2}
\hat{O} = \sum_{\boldsymbol{\sigma},\boldsymbol{\sigma^{\prime}}} V^{\sigma_1,\sigma^{\prime}_1}V^{\sigma_2,\sigma^{\prime}_2} \cdots V^{\sigma_{N-1},\sigma^{\prime}_{N-1}} V^{\sigma_N,\sigma^{\prime}_N} | \boldsymbol{\sigma} \rangle \langle\boldsymbol{\sigma^{\prime}} | ,
\end{align}
where we have omitted the sums over auxiliary indices as they can be recognized as matrix multiplications. At this point we note that, technically, a density matrix should be represented as an MPO instead of an MPS. However, the vectorization procedure allows the density matrix to be represented as an MPS and, as we shall see, the Liouvillian propagator to be represented as an MPO.

\subsubsection{Real time evolution}

To obtain the NESS, we target the solution of the master equation numerically given by
\begin{align}
\label{eq:solutionmps}
| \rho(\tau) \rangle = e^{\hat{W}\tau} | \rho(0) \rangle,
\end{align}
in the limit $\tau \rightarrow +\infty$, with $| \rho(\tau) \rangle$ being the density matrix of the state at time $t = \tau$, $| \rho(0) \rangle$ describing the density matrix of the initial state, and $\hat{W}$ being a vectorized form of the superoperator $\mathcal{W}$ in Eq.~\eqref{eq:supop}. As mentioned before, in this form, $\hat{W}$ corresponds to a square non-Hermitian matrix, while the density operators correspond to vectors in an extended Hilbert space.

The Liouville superoperator can be written as a sum of terms involving only two sites
\begin{align}
\label{twositeterms}
\hat{W} = \sum_{i = 1}^{N-1}\hat{W}_{i, i+1},
\end{align}  
given that the Hamiltonian involves only two-site terms and the Lindblad operators act locally. This structure allows one to introduce the so-called Trotter decomposition of the Liouville propagator. 

The first-order decomposition can be written as 
\begin{align}
\label{eq:firstordertrotter1}
e^{\hat{W}\tau} = \prod_{i=1}^{N-1} e^{\hat{W}_{i,i+1}\tau} + \mathcal{O}(\tau ^2).
\end{align}
The error introduced in this decomposition is due to the fact that nearest-neighbor Hamiltonian terms do not commute. However, next nearest neighbor Hamiltonian terms do commute, and this enables an even-odd decomposition of the Liouville propagator that can be carried out at the same time. In such a way, we can define 
\begin{align}
\label{eq:firstordertrotter2}
&\hat{O}_{\textrm{odd}} \coloneqq e^{\hat{W}_{1,2}\tau} \otimes \mathds{1} \otimes e^{\hat{W}_{3,4}\tau} \otimes \mathds{1} \otimes \cdots\,, \\
&\hat{O}_{\textrm{even}} \coloneqq \mathds{1} \otimes e^{\hat{W}_{2,3}\tau} \otimes \mathds{1} \otimes e^{\hat{W}_{4,5}\tau} \otimes \cdots\,,
\end{align}
such that $\hat{O}_{\textrm{odd}}$ and $\hat{O}_{\textrm{even}}$ can be applied at the same time $\tau$. One can notice that each of the $e^{\hat{W}_{i,i+1}\tau}$ acts on two sites so, in this form, the MPO structure is no longer present. To recover the MPO form, we need to decompose the operators in a way that preserves the locality attributed to the MPS. To do that, we can reshape the operators and apply SVD operations while keeping the maximum $\chi$ under control. To attain higher accuracy, instead of implementing the first-order decomposition as described, we use a higher order approximation; namely, the fourth-order Trotter-Suzuki decomposition given by
\begin{align}
\label{eq:fourthordertrotter1}
e^{\hat{W}\tau} = \mathcal{\hat{U}}(\tau_1)\mathcal{\hat{U}}(\tau_2)\mathcal{\hat{U}}(\tau_3)\mathcal{\hat{U}}(\tau_2)\mathcal{\hat{U}}(\tau_1) + \mathcal{O}(\tau^5),
\end{align}
with
\begin{align}
\label{eq:fourthordertrotter2}
\mathcal{\hat{U}}(\tau_i) = e^{\hat{W}_{\textrm{odd}}\tau_i / 2}e^{\hat{W}_{\textrm{even}}\tau_i }e^{\hat{W}_{\textrm{odd}}\tau_i / 2},
\end{align}
and
\begin{align}
\label{eq:fourthordertrotter3}
\tau_1 = \tau_2 = \frac{\tau}{4-4^{1/3}};\; \tau_3 = \tau - 2\tau_1 - 2\tau_2.
\end{align}

Once these MPOs are operated in the sequence shown in Eq.~\eqref{eq:fourthordertrotter1} on an initial state $| \rho(0) \rangle$, the MPS for $| \rho(\tau) \rangle$ is obtained. This procedure is done iteratively until the NESS is reached in light of Eq.~\eqref{eq:ness}, evaluating expectation values of observables after each time step. To contract the Liouville propagator in MPO form and the MPS at time $t$, we combine both methods presented in Ref.~\cite{Schollwock2011} to contract an MPS: SVD truncation and the variational approach. We find that convergence is achieved by providing the SVD-truncated state as an initial guess for the variational algorithm with only a few variational sweeps ($\approx$ 3-5). This approach provides better numerical results than using one of the two contraction methods on its own for a fixed value of $\chi$, albeit at a higher computational cost. We refer the reader to Refs.~\cite{Schollwock2011, Verstraete2008} for details on both contraction techniques.

The method described has two main sources of error. The first one is a truncation error due to the maximum value of the bond dimension $\chi$ used. In the specific case of simulations to reach nonequilibrium steady states, this error strongly depends on the system size $N$, the strength of the driving $\mu$, and the interaction parameter $\Delta$. The second source of error is related to the Trotter-Suzuki decomposition from Eq.~\eqref{eq:fourthordertrotter1}, which introduces an error of order $\mathcal{O}(M\tau^5)$ for the M-th time step. This error has also been found to depend linearly on the system size $N$ \cite{GobertDMRG2005}. In the case of NESS simulations, this error is not as important as the truncation error, given that the state does not change after the NESS is reached. In practice, in light of Eq.~\eqref{eq:homogeneous}, we apply enough time steps such that the standard deviation of the expectation value of the current operator averaged over all sites becomes very small ($\approx 0.5\%$).

\section{Linear response and error analysis from boundary driving configurations}
\label{ap:error}

\begin{figure}[!t]
\fontsize{13}{10}\selectfont 
\centering
\includegraphics[width=\columnwidth]{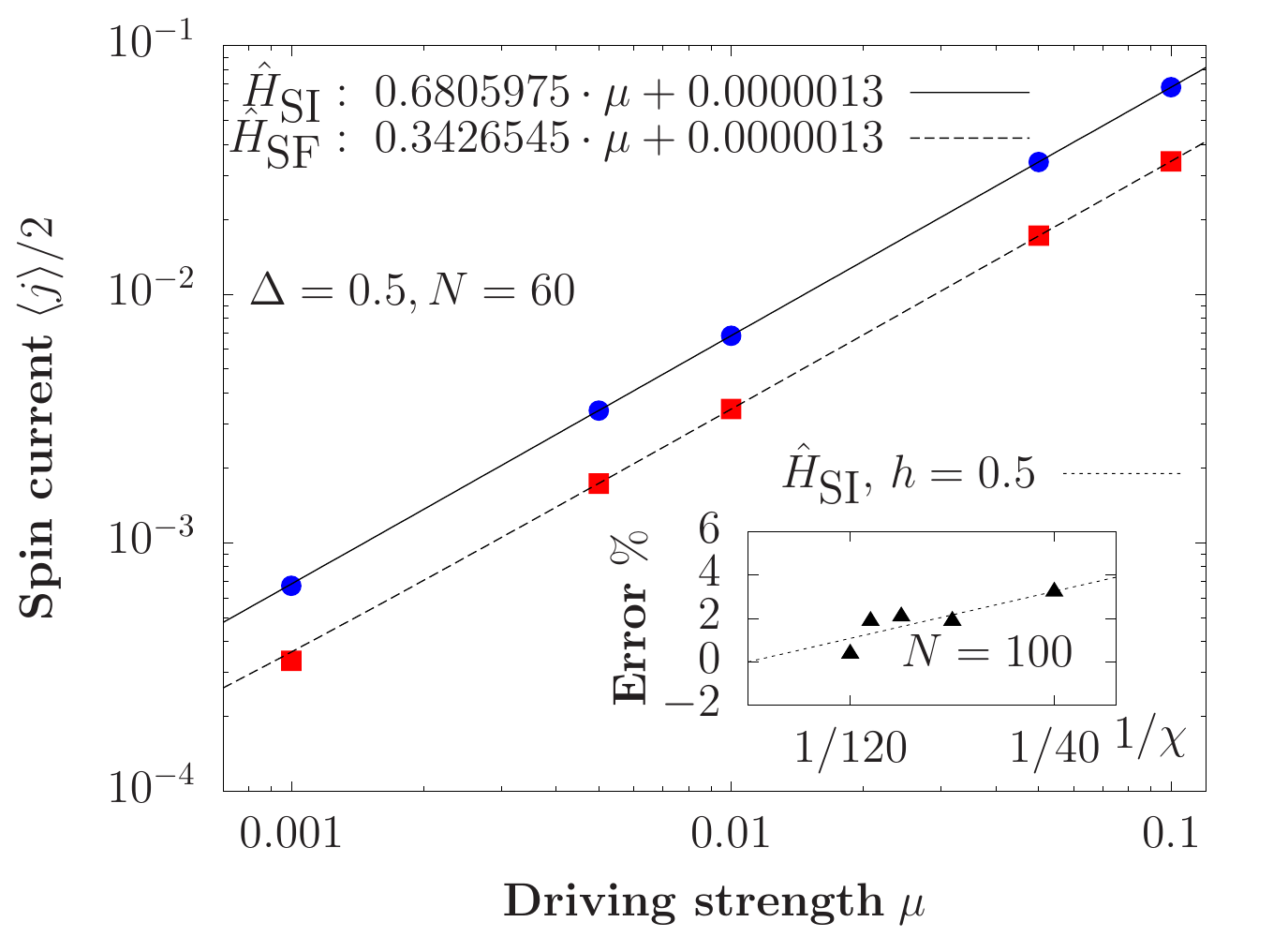}
\caption{Expectation value of the current in the NESS as a function of driving strength. (inset) Truncation error in the tMPS method versus the bond dimension $\chi$ for the largest system size we simulated for the $\hat{H}_{\textrm{SI}}$ model with $h = 0.5$. }
\label{fig:aperr}
\end{figure}

We have compared transport properties of different models using nonequilibrium configurations and linear response theory, in this section we demonstrate that the nonequilibrium transport calculations are within linear response regime.  Figure~\ref{fig:aperr} shows that the magnitude of the spin current depends linearly on the driving strength for values well above those used in our simulations. This implies that the transport properties in our systems depend linearly on $\mu$, $\langle \hat{j} \rangle \propto \mu$, and can be well-captured by linear response theory. For $\mu = 0$, the fit shown in Fig.~\ref{fig:aperr} is very close to zero, as no boundary driving implies no excitations propagating through the chain.

We analyzed the truncation error (induced by using a finite value of $\chi$) by studying the expectation value of the current operator [Eq.~\eqref{eq:spincurrent}] for the largest system size we simulated at fixed $\mu$ for different values of $\chi$. We then selected a value of $\chi$ that introduces a small tolerable error in our simulations. In the inset of Fig.~\ref{fig:aperr}, we show the error defined as $|\langle j(\chi) \rangle - \langle j(\infty) \rangle | / \langle j(\infty) \rangle \times 100$, where $\langle j(\infty) \rangle$ is an extrapolated value of the current, as a function of the bond dimension $\chi$. The scaling of the bond dimension suggests convergence for $\chi \rightarrow \infty$, as expected. In our calculations, we used $\chi = 100$ which results in an error due to the truncation that is $\lesssim 2\%$.

\section{Transport in the noninteracting regime}
\label{ap:nonint}

\subsection{NESS}
\label{ap:nonintbound}

Here we discuss the results when $\Delta=0$ in the models studied in the main text, i.e., in their noninteracting limit. In this limit, all these models are (trivially) integrable, and one can use the approach proposed in Ref.~\cite{DephasingMarko2013} to solve large system sizes at a low computational cost. Within this approach, a perturbative expansion is used to obtain the exact form of the nonequilibrium steady state by solving an equation of the Lyapunov type for any value of the boundary driving strength $\mu$ (we use $\mu = 1$). In the noninteracting limit, the dependence of the expectation value of the local magnetization and the spin current on $\mu$ is always linear. This is in contrast with the interacting case, which shows a nonlinear dependence for sufficiently strong boundary driving \cite{NegDiffCondBenenti2009}. 

\begin{figure}[!t]
\fontsize{13}{10}\selectfont 
\centering
\includegraphics[width=0.95\columnwidth]{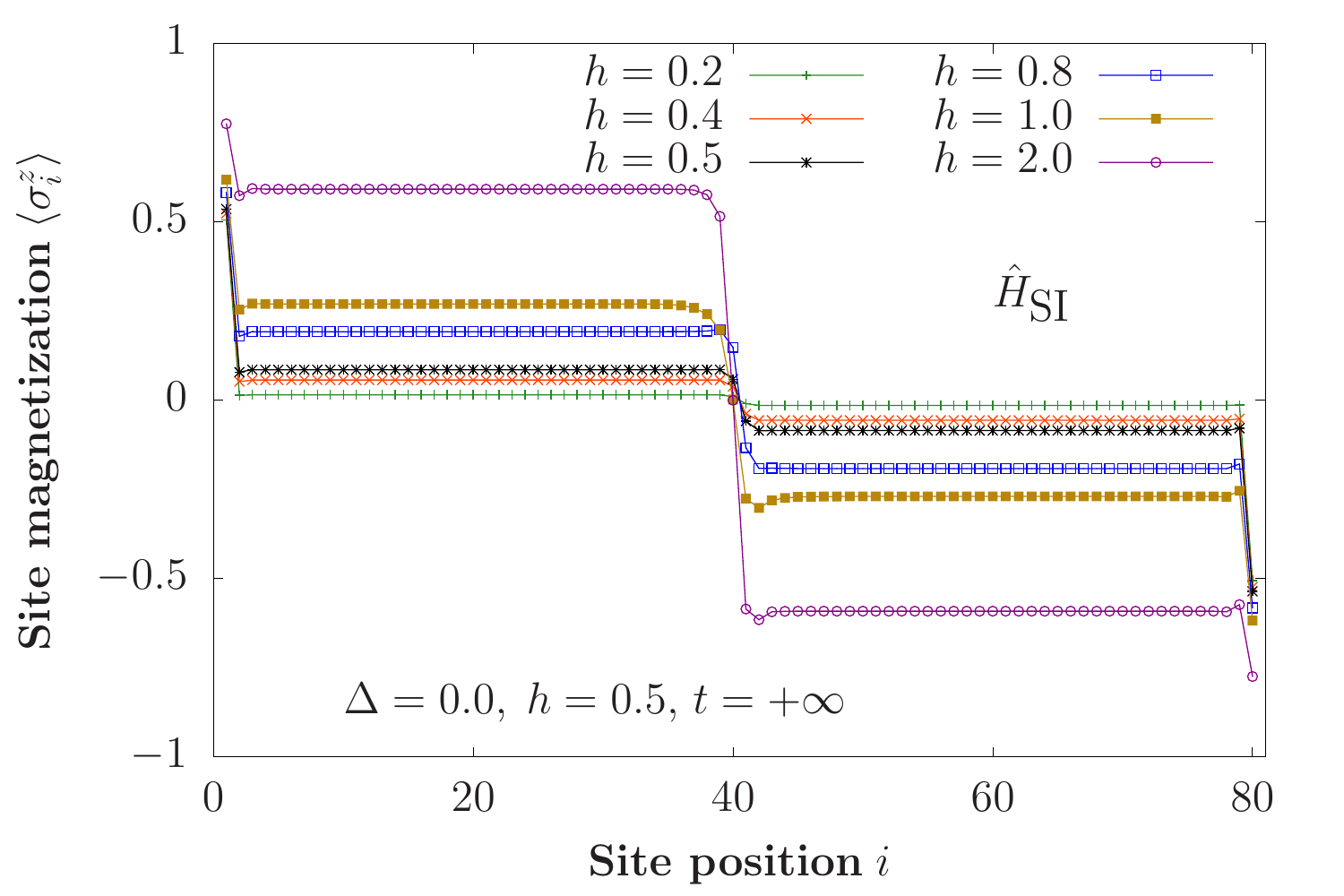}
\caption{Magnetization profiles in the NESS of the noninteracting limit of the $\hat{H}_{\textrm{SI}}$ model ($\Delta = 0$) in the presence of a single magnetic impurity with different strengths $h$. See Fig.~\ref{fig:6} for results when $\Delta \neq 0$. The profiles were obtained with $N = 80$, $\gamma = 1.0$, and $\mu = 1.0$.}
\label{fig:ap1}
\end{figure}

In Fig.~\ref{fig:ap1}, we show the magnetization profile of the NESS for the $\hat{H}_{\textrm{SI}}$ model with $\Delta = 0$ for different values of the impurity strength. The magnetization profiles are qualitatively similar to those in the interacting case depicted in Fig.~\ref{fig:6}, with the exception of the magnetization in the close vicinity of the impurity. For the interacting case, the magnetization profile in the vicinity of the impurity is somewhat smoother, while the noninteracting case exhibits an abrupt step.

\begin{figure}[!t]
\fontsize{13}{10}\selectfont 
\centering
\includegraphics[width=\columnwidth]{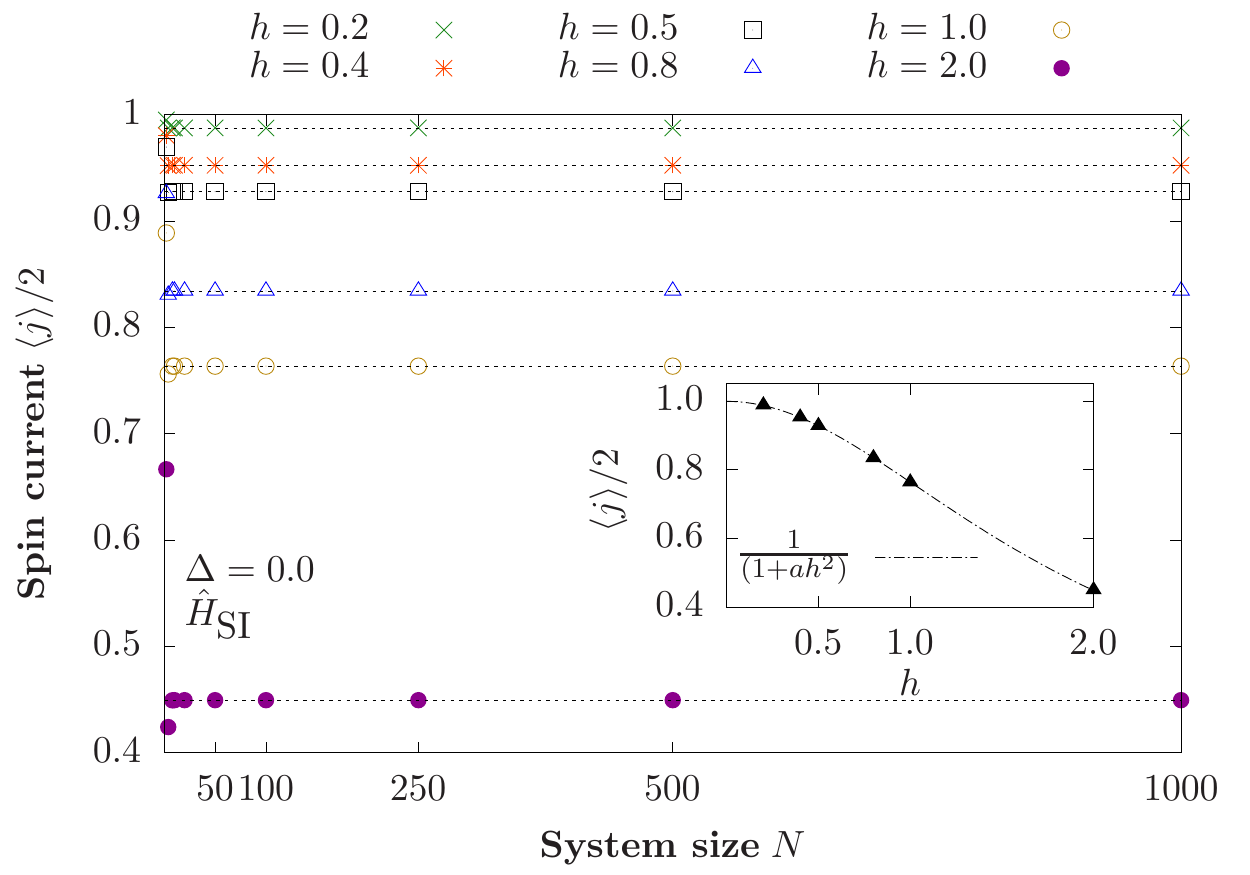}
\caption{Scaling of the expectation value of the current operator in the nonequilibrium steady state of the $\hat{H}_{\textrm{SI}}$ model with $\Delta = 0$ as a function of system size ($N=4,\cdots,1000$), for different values of $h$. The driving parameters are $\gamma = 1.0$ and $\mu = 1.0$.}
\label{fig:ap2}
\end{figure}

In Fig.~\ref{fig:ap2} we show the expectation value of the spin current operator in the NESS $\langle \hat{j} \rangle$ as a function of the chain sizes. One can see that, for sufficiently large system sizes, $\langle \hat{j} \rangle$ becomes independent on $N$, in analogy to the results for the interacting case in Fig.~\ref{fig:7}. The absolute value of $\langle \hat{j} \rangle$ decreases with increasing the strength of the impurity as $1 / (1 + ah^2)$ (see the inset in Fig.~\ref{fig:ap2}). This functional form is obtained from the transmission probability of free particles through a barrier at high temperatures \cite{ryndyk2016nano}. This is the functional form used in the fit reported in the inset in Fig.~\ref{fig:7} for the interacting case.

\begin{figure}[!b]
\fontsize{13}{10}\selectfont 
\centering
\includegraphics[width=\columnwidth]{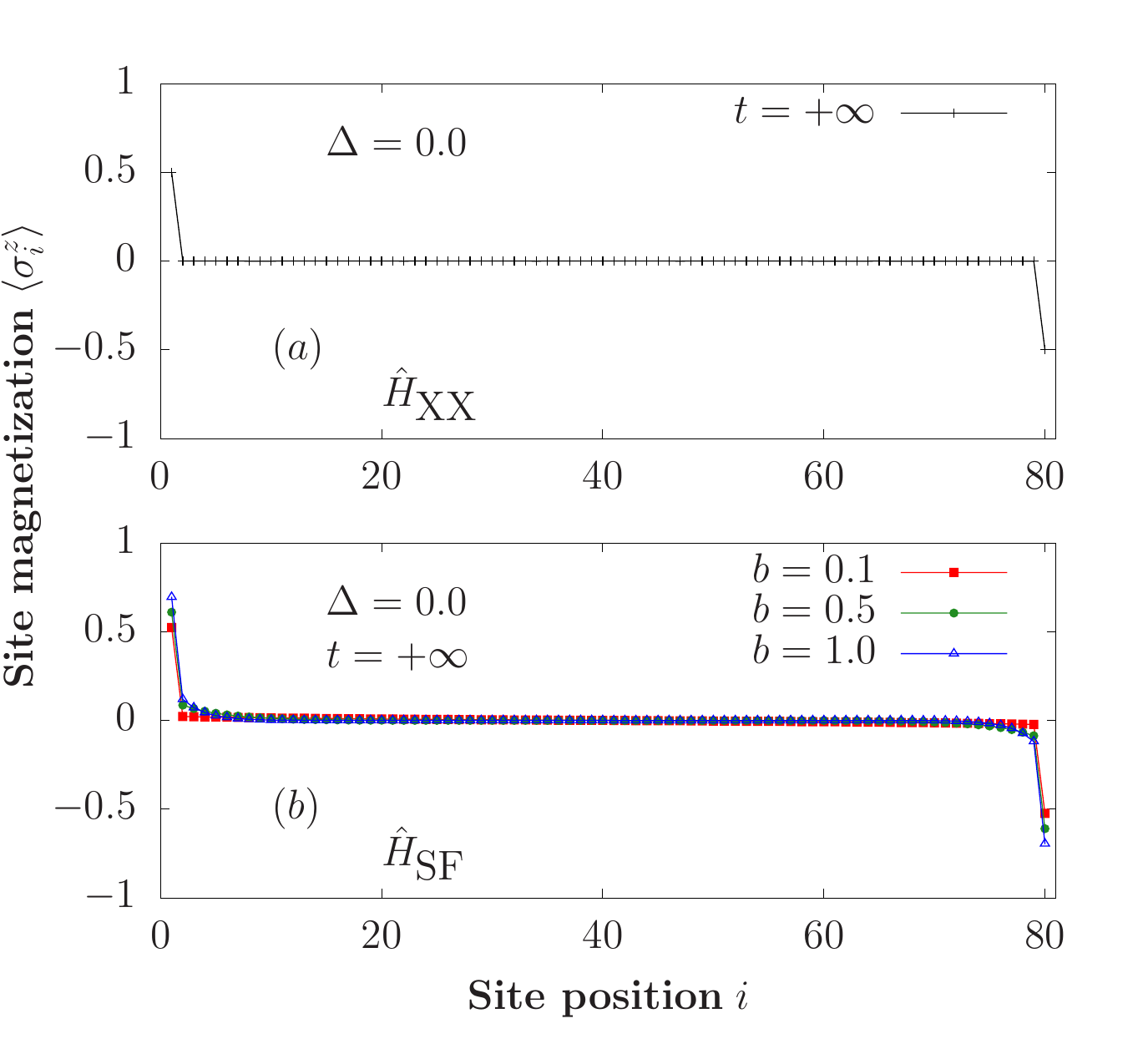}
\caption{(a) Magnetization profile of the nonequilibrium steady state for the noninteracting model and (b) for the noninteracting model in the presence of a staggered magnetic field with different values of $b$. The driving parameters are $\gamma = 1.0$ and $\mu = 1.0$.}
\label{fig:ap3}
\end{figure}

\begin{figure}[!t]
\fontsize{13}{10}\selectfont 
\centering
\includegraphics[width=\columnwidth]{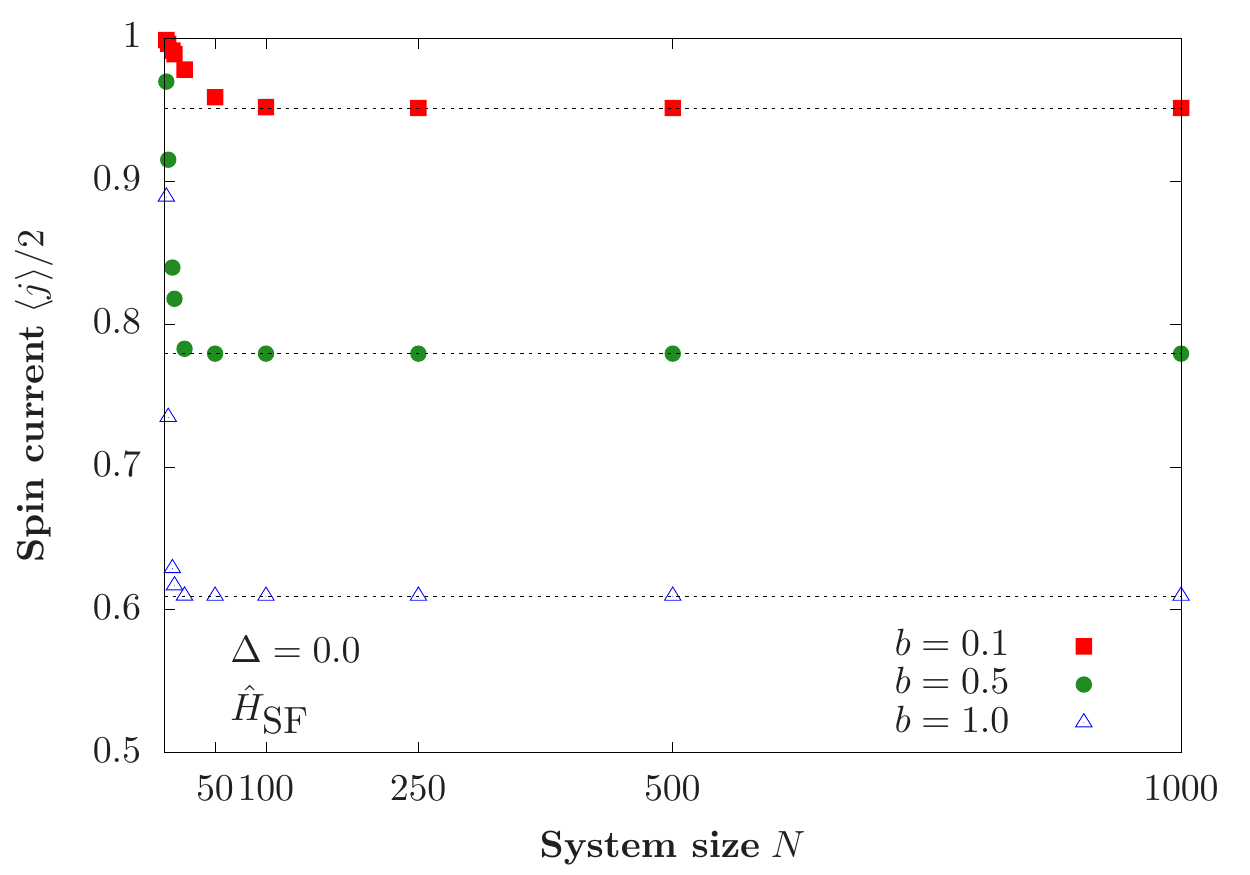}
\caption{Scaling of the expectation value of the current operator in the nonequilibrium steady state as a function of system size ($N=4,\cdots,1000$) for the noninteracting model in the presence of a staggered magnetic field. The driving parameters are $\gamma = 1.0$ and $\mu = 1.0$.}
\label{fig:ap4}
\end{figure}

In contrast to the results for the interacting and noninteracting $\hat{H}_{\textrm{SI}}$ models, the results for the magnetization profiles of the interacting and noninteracting $\hat{H}_{\textrm{SF}}$ models are fundamentally different. In Fig~\ref{fig:ap3}(a), we show the magnetization profile of the noninteracting XX model (the XXZ model for $\Delta=0$), to accentuate its similarity with the results for the noninteracting $\hat{H}_{\textrm{SF}}$ model reported in Fig~\ref{fig:ap3}(b), which are in stark contrast to the profiles for the interacting case reported in Fig.~\ref{fig:8}. In the noninteracting regime, the characteristic linear ramp-like profile observed for interacting systems that display incoherent transport (Fig.~\ref{fig:8}) is no longer present [Fig~\ref{fig:ap3}(b)]. As expected for systems with coherent transport, Fig.~\ref{fig:ap4} shows that the expectation value of the spin current operator $\langle \hat{j} \rangle$ in the noninteracting $\hat{H}_{\textrm{SF}}$ model becomes independent of $N$ for sufficiently large system sizes.

\subsection{Linear response theory}
\label{ap:nonintlinear}

For translational-invariant models in the noninteracting regime, such as the $\hat{H}_{\textrm{XX}}$ model, the properties of the total current operator $\hat{J}$ [Eq.~\eqref{eq:totalcurrent}] can be calculated analytically. In the free-fermion representation~\cite{Cazalilla:2011}, the eigenstates of the single-particle Hamiltonian are plane waves
\begin{align}
\label{eq:planewaves}
\ket{m} = \frac{1}{\sqrt{N}}\sum_{j}e^{ik_mj}c^{\dagger}_j\ket{0},
\end{align}
where $\ket{m}$ is the $m$-th eigenstate, with energy $\epsilon_m = -4\alpha\cos{(k_m)}$, $c^{\dagger}_j$ is the fermionic creation operator on site $j$, $\ket{0}$ is the vacuum state, and $k_m = 2\pi m / N$ with $m = -L/2 + 1, \cdots, L/2$. From this, the matrix elements of the total current operator are given by $|J_{nm}|^{2} = [4\alpha\sin{(k_m)}]^2\delta_{nm}$, i.e., the total current operator is diagonal in the energy eigenbasis. This implies that the second term in Eq.~\eqref{eq:drude1} is zero, and we obtain $D_N / (\langle -\hat{T} \rangle / N) = 1$ for any value of $N$.

\begin{figure}[!t]
\fontsize{13}{10}\selectfont 
\centering
\includegraphics[width=0.90\columnwidth]{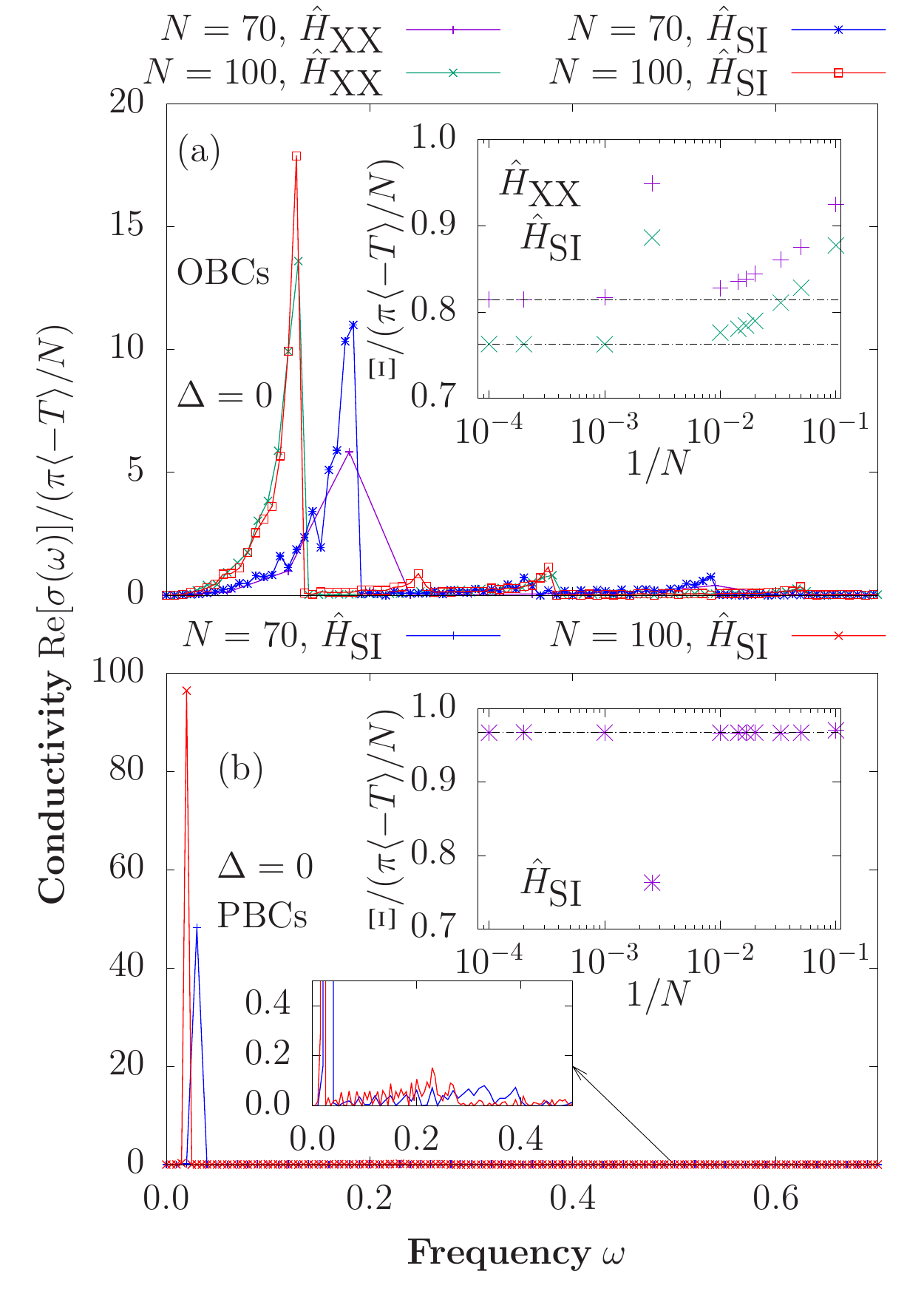}
\caption{Finite frequency part of the conductivity $\textrm{Re}[\sigma_N(\omega)]$ in noninteracting systems ($\Delta=0$). (a) Noninteracting limits of the $\hat{H}_{\textrm{XXZ}}$ and $\hat{H}_{\textrm{SI}}$ models with open boundary conditions. The inset in (a) shows the weight of the lowest frequency peak as a function of the system size. (b) and its bottom inset, Noninteracting limit of the $\hat{H}_\text{SI}$ model with periodic boundary conditions. The top inset in (b) shows the weight of the lowest frequency peak as a function of the system size. The results were obtained at very high temperature $\beta = 0.001$.}
\label{fig:ap5}
\end{figure}

On the other hand, as discussed in Sec.~\ref{sec:linearresponse}, chains with open boundary conditions have $D_N=\bar D_N=0$ irrespective of the presence or absence of interactions~\cite{RigolShastry2008}. Remarkably, $D_N=\bar D_N=0$ for the single impurity model in the noninteracting limit even in systems with periodic boundary conditions. This is the case because the impurity breaks the degeneracies between the single-particle $k$ and $-k$ eigenkets present in the translationally invariant case. Since the noninteracting limits of the XXZ and single impurity models are trivially integrable and must exhibit coherent transport, it is already apparent in this limit that the finite frequency part of Eq.~\eqref{eq:kuboformula} needs to be studied to compute the Drude weight~\cite{RigolShastry2008}.

In Fig.~\ref{fig:ap5}(a), we show the finite-frequency part of the conductivity in the noninteracting limit of the $\hat{H}_{\textrm{XXZ}}$ and $\hat{H}_{\textrm{SI}}$ models with open boundary conditions.  Since $D_{N} = 0$ in both cases, the sum rule in Eq.~\eqref{eq:sumrule} is fully accounted for by the finite-frequency part of the conductivity. Figure~\ref{fig:ap5}(a) shows that, with increasing system size in both models, the peaks present at finite frequency move toward $\omega=0$ (their frequency is $\omega \propto 1/N$~\cite{RigolShastry2008}) and become sharper. The weight of the peaks converge to a nonvanishing size-independent value with increasing system size. The inset in Fig.~\ref{fig:ap5}(a) shows the weight $\Xi$ of the lowest frequency peak (located at $\omega \approx 4\pi/N$) as a function of system size (the weight is two times the area under the peak). These results show that, in the thermodynamic limit, the systems develop a peak at $\omega = 0$ stemming from the collapse of peaks present at finite frequencies in finite systems. The weight of such a zero-frequency peak in systems with open boundary conditions is exactly the Drude weight predicted in systems with periodic boundary conditions~\cite{RigolShastry2008}.

Figure~\ref{fig:ap5}(b), and its bottom inset, show the finite-frequency part of the conductivity in the noninteracting limit of the $\hat{H}_{\textrm{SI}}$ model with periodic boundary conditions. The top inset in Fig.~\ref{fig:ap5}(b) shows the scaling of the weight of the lowest frequency peak as a function of system size. The same conclusions drawn for chains with open boundary conditions apply for chains with periodic boundary conditions. The lowest frequency peak, however, is much closer to $\omega=0$ and is much sharper in chains with periodic boundary conditions. Also, the weight of the lowest frequency peak is higher for periodic boundary conditions [see the top inset in Fig.~\ref{fig:ap5}(b) vs the inset in Fig.~\ref{fig:ap5}(a)]. In the thermodynamic limit, the lowest frequency peak almost accounts for the Drude weight in chains with periodic boundary conditions. 

The results for noninteracting systems discussed here, given the trivial nature of their coherent transport, highlight the subtleties discussed in Sec.~\ref{sec:linearresponse} when dealing with Kubo's linear response theory in systems without translational invariance. One needs to study the finite-frequency response in such systems in order to be able to determine whether transport is coherent or incoherent.

\bibliography{bibliography.bib}

\end{document}